\documentclass[aps,pra,twocolumn,showkeys]{revtex4-1}
\usepackage[utf8]{inputenc}
\usepackage{amsmath,amsthm,paralist}
\usepackage{amsfonts}
\usepackage{amssymb}
\usepackage{tikz}
\usepackage{times}
\usepackage{algcompatible}
\usepackage{algorithm} 
\usepackage[english]{babel}
\usepackage[justification=justified]{caption}
\usepackage{subcaption}

\newcommand{\nix}[1]{}
\newcommand{\etal}{{\em et al. }}
\newcommand{\ket}[1]{|#1\rangle}
\newcommand{\bra}[1]{\langle#1|}

\newcommand{\syn}{\text{syndrome}}
\newcommand{\F}{\mathbb{F}}

\newtheorem{theorem}{Theorem}
\newtheorem{proposition}[theorem]{Proposition}

\newtheorem{remark}[theorem]{Remark}

\usepackage[hidelinks,colorlinks=true,urlcolor= blue,linkcolor = blue,citecolor = red]{hyperref}

   \makeatletter
     \renewcommand\@make@capt@title[2]{%
      \@ifx@empty\float@link{\@firstofone}{\expandafter\href\expandafter{\float@link}}%
       {\textbf{#1}}\@caption@fignum@sep#2\quad}%
\makeatother

\begin{document}
\title{Efficiently decoding the 3D toric codes and welded codes on cubic lattices}
\author{Abhishek Kulkarni}
\author{Pradeep Kiran Sarvepalli}
\affiliation{Department of Electrical Engineering, Indian Institute of Technology Madras, Chennai 600 036, India}
\date{August 9, 2018}

\begin{abstract}
The recent years have seen a growing interest in quantum codes in three dimensions (3D). One of the earliest proposed 3D quantum codes is the 3D toric code. It has been shown that 3D color codes can be mapped to 3D toric codes. The 3D toric code on cubic lattice is also a building block for the welded code which has highest energy barrier to date. Although well known, the performance of the 3D toric code has not been studied extensively. In this paper, we propose efficient decoding algorithms for the 3D toric code on a cubic lattice with and without boundaries and report their performance for various quantum channels. We observe a threshold of  $\gtrsim 12\%$ for the bit flip errors, $\approx 3\%$ for phase flip errors and $ 24.8\%$ for erasure channel. We also study the performance of the welded 3D toric code on the quantum erasure channel. We did not observe a threshold for the welded code over the  erasure channel.
\end{abstract}
\maketitle

\section{Introduction} 

Three-dimensional (3D) toric codes are an important class of  topological codes. 
Kubica \etal   showed that  3D color codes can be mapped to copies of  3D toric codes \cite{kubica15}. 
Aloshious \etal showed  3D color codes can be projected onto 3D toric codes \cite{aloshious16}. 
These results highlight the importance of 3D toric codes. 
For instance,  3D color codes can be decoded via  3D toric codes. 
The computational power of 3D color codes becomes portable to 3D toric codes through code switching. 
The 3D toric code on the square lattice (with boundaries) is a building block for the welded code proposed by 
Michnicki \cite{michnicki14}. 
The welded code is particularly interesting because it has the highest known energy barrier to date. 
Further, Siva \etal showed that the memory time of welded code is doubly exponential in inverse temperature \cite{siva17}.

In some realistic quantum channels, there is an asymmetry in the bit flip and dephasing errors \cite{ioffe08}. 
Considerable benefits can be obtained by taking such asymmetry into account \cite{brooks13}.
These results suggest that the asymmetric error correcting capability of the 3D toric code for bit flip and phase flip errors 
could  be exploited in  quantum channels  where the bit flip and phase flip errors occur with different probabilities.

{\em In this paper we are interested in  studying the performance of 3D toric codes over various quantum channels.} 
Although  3D toric codes are among the  earliest known quantum codes, their performance has not been studied extensively.

Our work on the 3D toric code was also motivated in the context of the welded code, which is 
 a 3D quantum code built from many copies of the  3D toric code on a cubic lattice \cite{michnicki14}. 
 While the welded code was proposed as a candidate for quantum memory, it is not self-correcting. 
{\em This motivates another problem we study in this paper: efficiently decoding the welded code.}  
As the welded code is composed of 3D toric codes, it is  natural to try to decode the welded code by decoding the component 3D toric codes. 
This is another reason why we seek to decode the 3D toric code efficiently. 

Our contributions are as follows:
\begin{compactenum}[i)]
\item First we propose efficient decoders for the 3D toric code over the bit flip channel and the phase flip channel. 
With our decoding algorithms we obtain a threshold of  about $ 12\%$ for  bit flip channel, and $\approx3\%$ for the phase flip channel. 
These results build on the work of Dennis \etal \cite{dennis02} and Wang \etal \cite{wang10}.
After the completion of this work we came to know of a result by Duivenvoorden \etal \cite{duivenvoorden17} who proposed a renormalization decoder which gives a threshold of 17.2\% for the bit flip channel. 
\item Secondly, we propose a decoding algorithm for the 3D toric code over the quantum erasure channel  (QEC).
This extends the work of Delfosse \etal \cite{delfosse17} on 2D toric codes to 3D. 
We obtained a threshold of  $24.8\%$ for the quantum erasure channel. 
This is very close to the bond percolation threshold  of the cubic lattice \cite{wilke83}
suggesting that the proposed algorithm's performance is almost optimal. 

\item Thirdly, we propose an efficient decoder for the welded toric codes over QEC.
For the welded code we did not observe a threshold over the quantum erasure channel. 
This is in agreement with the claim made in \cite{arxiv_michnicki14} that the welded toric code does not have a phase transition.
\end{compactenum}

The 3D toric code is a Calderbank-Shor-Steane (CSS)  code \cite{calderbank98} in which $X$ and $Z$ errors can be corrected independently. 
However, as mentioned earlier, it has asymmetric error correcting capabilities for the bit flip and phase flip errors,
hence, decoding them independently entails the use of different decoders. 

From \cite{dennis02} it is implicit  that a combination of the matching decoder used for 2D toric codes and a generalization of the 
Toom's rule 
can lead to a  decoder for the 3D toric code on the cubic lattice for independent Pauli errors.
To elaborate, errors on the 3D toric code share some aspects with the 2D toric codes on the one hand and the 4D toric codes on the other. 

The phase errors are string-like and we use a decoder based on matching. 
In case of 3D toric code with boundaries, the matching algorithm must 
 be modified to account for them. We adapt the algorithm proposed in \cite{wang10} for 2D toric codes. This decoder is applicable to all 3D toric codes for phase flip errors.  

The bit flip errors are like surfaces. We use a local decoder based on the Toom's rule  for classical 2D memories. 
Cellular automata  decoders based on this rule have been proposed for the 4D toric code \cite{dennis02,pastawski11} and studied in
\cite{breuckmann16}. 
Our decoder is  an adaptation of Toom's rule to 3D in the presence of  boundaries. 
While our decoder is inspired by Toom's rule, as are the 4D decoders, it is deterministic unlike \cite{dennis02} and 
uses multiple rules unlike \cite{pastawski11}.
It is also capable of correcting errors which are not corrected by a straightforward adaption of the Toom's rule.

We also look at quantum erasure channel  which models the situation where qubits are lost or leaked.  
There are also multiple other physical scenarios where errors can be modeled by an erasure channel \cite{grassl97}. 
Classically, the erasure channel is studied extensively, not only because it is analytically more tractable, but, also because
of the insights it provides.

In the recent years, many researchers have turned their attention to the quantum erasure channel \cite{kudekar16,lloyd17,delfosse16,delfosse13}.
Delfosse \etal  proposed a  maximum likelihood decoder  for surface codes over the erasure channel \cite{delfosse17}. 
We provide a linear algebraic perspective on this decoder which could be of independent interest. 
We build upon this decoder and 
propose a decoding algorithm for 3D toric codes over the QEC.
The erasure decoding problem can also be reduced to decoding the bit flip and phase flip errors separately. 
For the correction of phase flip errors we use the approach proposed in \cite{delfosse17}. 
However, our implementation of the decoder takes a slightly different perspective.
In case of bit flip errors occurring under QEC, we propose a different algorithm.

We propose a decoding algorithm for the welded code using the 3D toric code decoder as a component.
The 3D toric code decoder cannot be used as it is because the 3D toric codes constituting the welded code are not independent but share some qubits. 
Our algorithm appropriately decouples them and decodes the welded code.
We did not observe a threshold for the welded codes.

The rest of the paper is organized as follows. 
We review the necessary background in Section~\ref{sec:bg}. In Section~\ref{sec:3dtc}, we present the decoders for the 3D toric code for the bit flip and phase flip errors.
In Section~\ref{sec:3dtc-qec} we propose a decoder for the 3D toric code over the quantum erasure channel. 
In Section~\ref{sec:welded} we study the performance of the welded codes over QEC. 
Finally, we conclude in Section~\ref{sec:conc} with a brief discussion on scope for future work. 

\section{Background} \label{sec:bg}

In this section we give a self contained review of 3D toric codes \cite{castelnovo08} and welded codes \cite{michnicki14}. 
We assume the readers are familiar with stabilizer codes \cite{calderbank98,gottesman97}.

\subsection{3D toric codes} Consider a (cubic) lattice $\Lambda$ in 3D. 
Qubits are placed on edges of $\Lambda$ and
for each vertex $v$ we define an $X$ type operator called the vertex operator
\begin{eqnarray}
S_v^X  = \underset{e\in \delta v}{\prod} X_e,\label{eq:vertex-stab}
\end{eqnarray}
where $\delta v$ is collection  of edges incident on $v$.
For each face $f$ we  define a $Z $ type operator called the plaquette operator
\begin{eqnarray}
S_f^Z  = \underset{e \in \partial f}{\prod} Z_e,\label{eq:face-op}
\end{eqnarray}
where $\partial{f}$ is collection of edges in the boundary of $f$.
The 3D toric code defined on $\Lambda$ is the stabilizer code whose stabilizer is generated by $S_v^X$ and $B_f^Z$ where $v$ and $f$ run over all vertices and faces of $\Lambda$ respectively.

\begin{figure}
        \begin{subfigure}[b]{0.2\textwidth}
                \centering
               \includegraphics[scale=1.2]{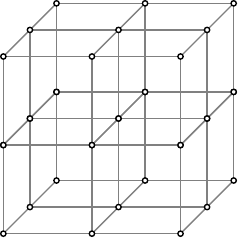}
				\caption{} 
				\label{fig:3DToricCode}
        \end{subfigure}
        \begin{subfigure}[b]{0.2\textwidth}
                \centering
               	\includegraphics[scale=.8]{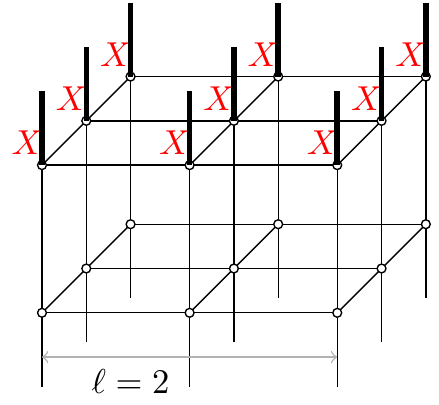}
			   	\caption{}
			   	\label{fig:solidcodeWithZ}
        \end{subfigure}
        \begin{subfigure}[b]{0.2\textwidth}
                \centering
               	\includegraphics[scale=0.8]{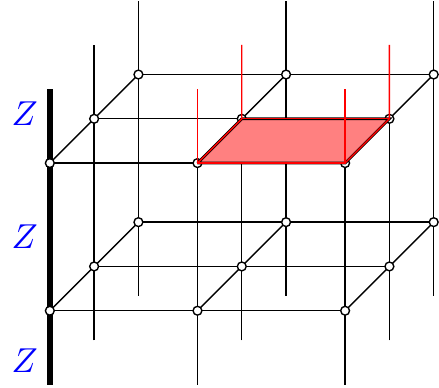}
				\caption{}
				\label{fig:solidcodeWithX}
        \end{subfigure}
        \begin{subfigure}[b]{0.2\textwidth}
                \centering
               	\includegraphics{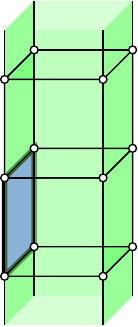}
				\caption{}
				\label{fig:cubeDependency}
        \end{subfigure}
		\begin{subfigure}[b]{0.2\textwidth}
                \centering
               	\includegraphics[scale=0.9]{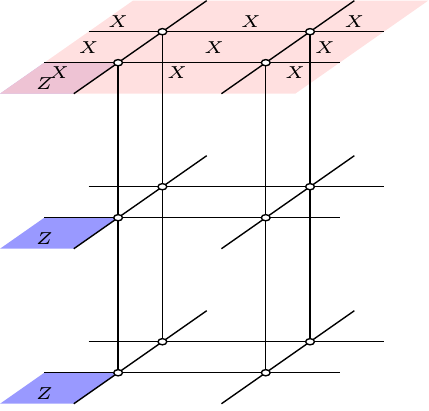}
				\caption{}
				\label{fig:dualcode}
        \end{subfigure}
        \captionsetup{justification=raggedright,singlelinecheck=false}
        \caption{(a)  3D toric code with periodic boundary. (b)  Solid  code i.e., 3D toric code with boundaries; also shown is 
        $\overline{X}$, the logical $X$ operator.  (c) A dependent horizontal face operator; also shown is $\overline{Z}$, the logical $Z$  operator.  (d)  Dependency among  vertical face operators in a stack of cubes. (e) Solid code in the dual lattice. Encoded operators  $\overline{X}$ (in red) and $\overline{Z}$ (in blue).}
 \end{figure}

Consider a cubic lattice in 3D, as in  Fig.~\ref{fig:3DToricCode}.
Under periodic boundary conditions all the vertex operators 
are of weight six and all face operators are of weight four.
The 3D toric code with periodic boundary conditions encodes three logical qubits \cite{castelnovo08}. 

We can introduce boundaries by allowing for half edges as shown in Fig.~\ref{fig:solidcodeWithZ}.
Unlike the 3D toric code on cubic lattice with periodic boundary conditions, here all the vertex operators 
are not of the same weight. 
Some vertex operators are of either weight four or five instead of six, as can be seen in Fig.~\ref{fig:solidcodeWithX}. 
Again in contrast to the periodic cubic lattice, all face operators are of not same weight. 
Some face operators are of weight 3 instead of 4, as can be seen in Fig.~\ref{fig:solidcodeWithX}. 
The collection of half edges on the top form a rough boundary.
Similarly, the  half edges on the bottom form another rough boundary.  
This code has been termed solid code in \cite{michnicki14} and encodes one logical qubit. 
For completeness we include this computation.

The total number of qubits, $n(\ell)$ in a solid code on a   cubic lattice of side $\ell$ as shown in Fig.~\ref{fig:3DToricCode}, is 
\begin{eqnarray}
n(\ell) =3\ell^3+5\ell^2+3\ell+1 \label{eq:noOfQubits}
\end{eqnarray}

All the vertex operators are independent. 
They are $\ell(\ell+1)^2$ in number. 
On the other hand, there are many dependencies among the face operators. 
All the operators on  faces  with half edges are independent. 
They are $4\ell(\ell+1)$ in number. 
Next observe that the operator associated to a horizontal face is dependent, see Fig.~\ref{fig:solidcodeWithX} for an illustration.
This leaves only the vertical face operators. 
There  are $2\ell(\ell+1)^2$ such operators. 

Consider all the vertical faces in one stack of cubes as shown in Fig.~\ref{fig:cubeDependency}. 
The product of all the respective face operators is identity which gives us one more dependency.

There are $\ell^2$ number of stack of cubes in solid hence $\ell^2$ number of such dependencies. 
Thus there are $2\ell(\ell+1)^2-\ell^2= 2\ell^3+3\ell^2+2\ell$ independent face operators. 
Totally, there are $s(\ell) = 3\ell^3+5\ell^2 +3\ell$ independent stabilizer generators, thus the solid code encodes 
$n(\ell)-s(\ell)$ logical qubits.

As mentioned earlier, the solid code is asymmetric in its error correcting capabilities. 
The $Z$ distance of the code is $\ell+1$, see Fig.~\ref{fig:solidcodeWithZ} while the $X$ distance is $(\ell+1)^2$, see
Fig.~\ref{fig:solidcodeWithX}.
Thus the solid code on a cubic lattice of  size $\ell$ is and $[[ n(\ell),1,\ell+1]]$ quantum code.

In correction of $X$ errors we use the dual lattice of solid code. 
The dual lattice  is obtained by one to one mapping  of vertices  to lattice cubes, edges to faces, faces to edges and lattice cubes to vertices. Two vertices are adjacent in the dual lattice, denoted $\Lambda^*$, if their preimages share 
a face in the original lattice.
In the dual lattice 
qubits are associated to faces,  $X$ stabilizers to cubes and $ Z$ stabilizers  to edges.
Dual lattice of the 3D toric code in Fig.~\ref{fig:solidcodeWithX} is shown in Fig. ~\ref{fig:dualcode}. 
The logical $X$ operator can be visualized as a surface in the dual lattice.

\subsection{Welded codes}

Motivated by the problem of quantum memory, 
Michnicki proposed a new type of
code construction for CSS codes called welding. 
Using this method he welded 3D toric codes to obtain the welded 3D toric code which has the largest known energy barrier.
(This code was termed welded solid code in \cite{michnicki14}.
Throughout this paper we shall refer to this code as the welded toric code or simply the welded code.)

We briefly review this construction and the 3D welded toric codes. 
We explain welding through a simple example. Let $S_1$ and $S_2$ stabilizers of two codes.
(These are 2D toric codes with boundaries.)
Let $\overline{X}_i$ and $\overline{Z}_i$ be the associated encoded operators for the $i$th code.

\begin{eqnarray*}
  S_1=\left[
  \begin{array}{ccccc}
   X & I & X & X & I \\
   I & X & X & I & X \\
   Z & Z & Z & I & I \\
   I & I & Z & Z & Z
  \end{array}\right];
  S_2=\left[
  \begin{array}{ccccc}
   X & I & X & X & I \\
   I & X & X & I & X \\
   Z & Z & Z & I & I \\
   I & I & Z & Z & Z \\
  \end{array} \right]\quad
\end{eqnarray*}
\begin{eqnarray*}
   \left[\begin{array}{c}
   \overline{X}_1\\
   \overline{Z}_1\\
   \end{array}
   \right]=\left[
  \begin{array}{ccccc}
   I & I & I & X & X \\
   Z & I & I & Z & I \\
  \end{array} \right]; 
   \left[\begin{array}{c}
   \overline{X}_2\\
   \overline{Z}_2\\
   \end{array}
   \right]=\left[
  \begin{array}{ccccc}
   I & I & I & X & X \\
   Z & I & I & Z & I \\
  \end{array} \right]\quad
\end{eqnarray*}

The first step of welding is to identify $w$ qubits from each code 
and consider them to be the same. 
Suppose that the fourth and fifth qubit of $S_1$ is identified as the first and second qubit of $S_2$ respectively.
After identification combine $S_1$ and $S_2$ as shown below.
\begin{align*}
 \left[ 
  \begin{array}{cccccccc}
   X & I & X & X & I \\
   I & X & X & I & X \\
   Z & Z & Z & I & I \\
   I & I & Z & Z & Z \\
   & & & X & I & X & X & I \\
   & & & I & X & X & I & X \\
   & & & Z & Z & Z & I & I \\
   & & & I & I & Z & Z & Z
  \end{array} \right]
\end{align*}

Now we can see that all generators of $S_1$ and $S_2$ do not form a commutative set and all of them cannot be included to form another stabilizer. 

Welding is a method to combine the  generators so that we obtain a commutative group. Two types of welding are possible. 
\begin{compactenum}[i)]
\item $Z$-weld: Extend the stabilizer groups $S_i$ by including $\overline{Z}_i$. This leads to new stabilizer codes with zero encoded qubits.
Then retain all the $X$ type stabilizer generators after extending them to act on all the qubits. Denote this set by $S_w^X$. 
Add all the $Z$ type generators 
which commute with the $X$ type generators. 
Then we include all
the $Z$ type stabilizers of $S_i$ which commute with $S_w^X$, after suitably extending them. 
Noncommuting 
$Z$ type operators are modified to  obtain a generator which commutes with the  all of $S_w^X$.
Finally the operator obtained by modifying the logical $Z$ operators is promoted to a logical operator.

\item $X$-weld: The converse of $Z$-weld, where 
$Z$ type generators are retained and $X$ type generators merged. 
\end{compactenum}
In the context of toric codes, $X$-weld and $Z$-weld are also referred to 
as smooth and rough welds, respectively. 

We illustrate the $Z$ weld with our running example. 
Adding the logical operators to 
$S_i$ and identifying the qubits gives the following set of operators. 
\begin{align}\label{S'matrix}
  S'&=\left[ 
  \begin{array}{cccccccc}
   X & I & X & X & I \\
   I & X & X & I & X \\
   Z & Z & Z & I & I \\
   I & I & Z & Z & Z \\
   & & & X & I & X & X & I \\
   & & & I & X & X & I & X \\
   & & & Z & Z & Z & I & I \\
   & & & I & I & Z & Z & Z \\
   Z & I & I & Z & I \\
   & & & Z & I & I & Z & I \\
  \end{array} \right]
\end{align}
Then keep all the $X$ type stabilizers and the $Z$ type stabilizers which commute with the $X$ type stabilizers of $S_1$ and $S_2$. 
The (extended) $Z$ type generators of $S_1$ and $S_2$ which commute with the $X$ type generators in $S'$ are
$Z  Z  Z \textcolor{red}{I} \textcolor{red}{I} I I I,\hspace{0.1cm} I I I \textcolor{red}{I} \textcolor{red}{I}  Z Z Z ,\hspace{0.1cm} I I Z \textcolor{red}{Z} \textcolor{red}{Z}  Z I I $.
The noncommuting stabilizer operators $I I Z \textcolor{red}{Z} \textcolor{red}{Z}  I I I$ and $I I I \textcolor{red}{Z} \textcolor{red}{Z}  Z I I$ get welded to form $I I Z \textcolor{red}{Z} \textcolor{red}{Z}  Z I I$.

The logical $Z$ operators of the welded code are  obtained by modifying the component logical $Z$ operators. 
More precisely, the 
$Z$ logical operators of the component codes, $ZI I \textcolor{red}{Z} \textcolor{red}{I}  I  I I$ and $I I I \textcolor{red}{Z} \textcolor{red}{I}  I  Z I$ are combined to form, $ZI I \textcolor{red}{Z} \textcolor{red}{I}  I  Z I$, the 
$Z$ logical operator of the welded code. 
Hence, the new code has the stabilizer $S_w$ given as

\begin{eqnarray}
  S_w&=\left[
  \begin{array}{cccccccc}
   X & I & X & \textcolor{red}{X} & \textcolor{red}{I} & I & I & I\\
   I & X & X & \textcolor{red}{I} & \textcolor{red}{X} & I & I & I\\
   I & I & I & \textcolor{red}{X} & \textcolor{red}{I} & X & X & I \\
   I & I & I & \textcolor{red}{I} & \textcolor{red}{X} & X & I & X \\
  I & I & Z & \textcolor{red}{Z} & \textcolor{red}{Z} & I & I &I\\
  I & I & I & \textcolor{red}{Z} & \textcolor{red}{Z} & Z & I & I\\
  I & I & Z & \textcolor{red}{Z} & \textcolor{red}{Z} & Z & I & I
  \end{array}\right]
\end{eqnarray}

Either $\overline{X}_1$ or $\overline{X}_2$ (appropriately extended) can be viewed as the  logical $X$ operator of the welded code. 
\begin{eqnarray}
   L_w=\left[
  \begin{array}{cccccccc}
   I & I & I & \textcolor{red}{X} & \textcolor{red}{X} & I & I & I\\
   Z & I & I & \textcolor{red}{Z} & \textcolor{red}{I} & I & Z & I\\
  \end{array} \right]\quad
\end{eqnarray}

As mentioned earlier, in  $Z$-weld,  $X$ type stabilizers do not change in weight.
The $Z$ type stabilizer generators which anticommute with the $X$ type generators
get welded and their weight increases, as does the weight of the logical $Z$ operators.

Welding as we described is mostly specific to  toric codes, details about  welding in general and additional technical conditions can be found in \cite{michnicki14,arxiv_michnicki14}.

Welding can be performed with multiple stabilizer codes. 
By repeatedly performing welding, the weight of the logical operators is increased for the welded code.
By properly choosing the number of times to weld the 3D toric codes, \cite{michnicki14} obtained codes with higher energy barrier. 
Higher energy barrier ensures increase in memory time.

\begin{figure}[H]
    \centering
    \includegraphics[scale=0.63]{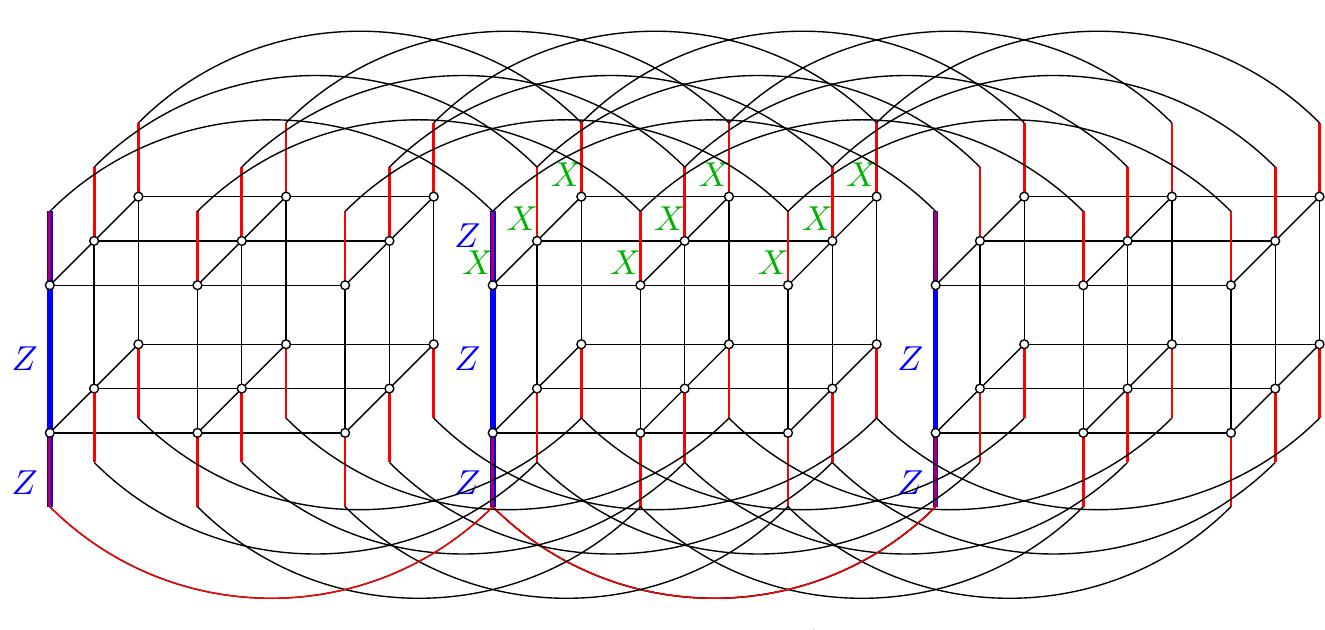}
    \caption{Welding of three  solid codes. 
    Welded qubits are shown in red.
    Also shown are the logical operators: $\overline{Z}$ (in blue) and $\overline{X}$ (in green). Welded code parameters are, $[[3n(\ell)-4(\ell+1)^2,1,O(3\ell)]]$, where $n(\ell)$ is given in Eq.~\eqref{eq:noOfQubits}.}
    \label{fig:WeldedCodeSimpleExample}
\end{figure}

We show two examples of welding of 3D solid code. In first example shown in Fig. \ref{fig:WeldedCodeSimpleExample} red lines represent welded qubits and curved line connecting them represent the weld, which means connected red qubits collectively represent the same qubit. One welded qubit is shown in red color for illustration. Rough weld is done here, which means $Z$-stabilizers at rough boundary get welded. It is also shown in Fig. \ref{fig:WeldedCodeSimpleExample} that 
the logical $Z$ operators (shown in blue color) of the component 3D toric code get welded  whereas $\overline{X}$ remains the unchanged  (shown in green color) up to the stabilizer of the welded code. 

In Fig. \ref{fig:WeldedCodeSimpleExample} three solid codes are welded together.  Welding is done in two places, top and bottom rough boundary. The number of boundary qubits in bottom and top rough boundary each is $(\ell + 1)^{2}$.
In bottom rough boundary we have $3(\ell+1)^{2}$ qubits before welding. After welding we have only $(\ell + 1)^{2}$. Similarly for top rough boundary. Therefore, total number of qubits after welding is $3n(\ell) -4(\ell+1)^{2}$.

\begin{figure}[H]
    \centering
    \includegraphics[scale=2]{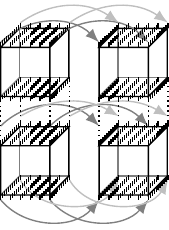}
    \caption{Welded code showing welding of four solid codes. 
    Welded code parameters are, $[[4n(\ell)-5(\ell+1)^2,1,O(4\ell)]]$, and $n(\ell)$ is given in
    Eq.~\eqref{eq:noOfQubits}.}
    \label{fig:WeldedCode}
\end{figure}

In the next example we show how solid codes stacked are above each other and welded. In Fig. \ref{fig:WeldedCode} all the dotted and curved lines represent the weld. Also, bottom and top rough boundary are welded together.  

Fig. \ref{fig:WeldedCode} shows solid codes welded along two directions. To get higher energy barrier welding solid codes stacked along two directions as in Fig. \ref{fig:WeldedCode} is not enough. We need to weld solid codes along all three directions,  $x$, $y$ and $z$.

Let $R$ be the number of solid codes stacked in each direction and then welded together by $Z$-weld. And if  each solid code is $\ell$ qubits wide and $\ell=O(R^2)$.  Total number of solid codes welded are $R^3$ and number of qubits in each solid code is $n(\ell)$. Next we will calculate number of qubits in welded code as done in example, Fig. \ref{fig:WeldedCodeSimpleExample}. Welded code with $R$ solid codes in each direction and will have $R+1$ number of places of weld. In bottom-most and top-most welds, $R^2$ solid codes are welded together and remaing $R-1$ welds, $2R^2$ solid codes are welded together. Total number of qubits in welded code is, 
\begin{eqnarray}
n_w(\ell)&=&R^3n(\ell) -(\ell +1)^{2}(2R^2-2) \nonumber\\
		&&\qquad\qquad - (R-1)(2R^2-1)(\ell+1)^{2} \nonumber\\
		&=&R^3n(\ell) -(\ell +1)^{2}(2R^3-R-1) \label{eq:welded-length}
\end{eqnarray}

In solid code the weight of $X$ and $Z$ stabilizers do not change with code length. In case of welded code, weight of welded $Z$ stabilizers changes; compared to solid code it increases by $R^2$. 
This means the weight of welded $Z$ stabilizers  increases with length. But weight of un-welded $Z$ and all $X$ stabilizer does not change with length.

In solid code (minimum) weights of $X$  and $Z$ logical operators are $\ell^{2}$ and $\ell$ respectively. As rough weld does not change the $X$ logical operator it's weight remains same in welded code. 
The weight of $Z$ logical operator changes to $O(R^3\ell) = O(\ell^{5/2})$.
The distance of the welded code is $\min \{ \ell^{2}, \ell^{5/2} \} =\ell^2$.
In \cite{arxiv_michnicki14} it was shown that welding of stabilizer codes with zero encoded qubits will lead to welded code with zero encoded qubits. Before welding we converted stabilizer codes to code with zero encoded qubits by including $\overline{Z}$ in stabilizer set. This  leads to a welded code with zero encoded qubits. Lastly,  the welded $\overline{Z}$ operator is promoted back to logical operator, giving one encoded qubit.
Hence welded code parameters will be, $[[R^3\ell^3,1,\ell^{2}]] $ where $R=O(\sqrt{\ell}))$.
In passing we note that it was shown in \cite{michnicki14} that the 3D welded code has an energy barrier $O(\ell)$.
If $n$ is the length of the code, then the welded code has parameters 
$[[n,1,O(n^{4/9})]]$ and its energy barrier is $O(n^{2/9})$.

\section{Decoding 3D toric code for  phase and bit flip errors}\label{sec:3dtc}

As the 3D toric code is a CSS code, we can decode the bit flip and phase flip errors separately. 
We focus on the 3D toric code with boundaries. 
Towards the end of the section we discuss how the decoder needs to be modified for the toric code with periodic boundary conditions. 
We end this section with simulation results of the toric code with boundaries.

\subsection{Correction of Phase errors} \label{ssec:3dtc-z-dec}

In this section, we show how to correct the phase errors. 
The structure of phase errors in the 3D toric codes is similar to that of 
$Z$ or $X$ errors on the 2D toric codes. 
So decoding schemes used for 2D toric codes can be  adapted for the 3D toric codes.
The easier case is when the toric code has periodic boundary conditions.
In this case every phase error violates an even number of vertex type checks and the errors can be 
identified with a collection of paths that terminate on these vertices whose checks are violated. 
We can then use the minimum weight perfect matching algorithm to find the most likely error as in case of the 2D toric codes. 

\begin{figure}
\begin{subfigure}[b]{0.2\textwidth}
                \centering
               	\includegraphics[scale=2.75]{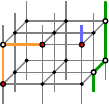}
\caption{}
\label{fig:singleZerrors}
        \end{subfigure}
        \begin{subfigure}[b]{0.2\textwidth}
                \centering
				\includegraphics[scale=1.25]{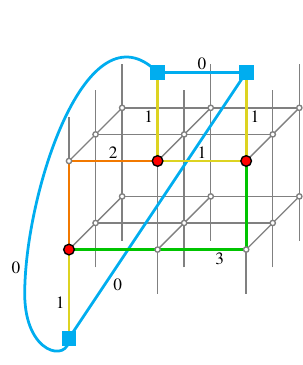}
				\caption{}
				\label{fig:completeGraph}
        \end{subfigure}
        \captionsetup{justification=raggedright,singlelinecheck=false}
        \caption{(a) Shows some phase error patterns. Errors are shown in color  and corresponding nonzero syndromes by  filled circles.   (b) Creating an  auxiliary graph $\mathfrak{K}$ (in bold) for the error 							pattern in Fig.~\ref{fig:singleZerrors}. Vertices of $\mathfrak{K}$ are the nonzero syndrome nodes of (a) and boundary nodes for each nonzero syndrome. Minimum weight perfect matching  algorithm is run on this graph. }
\end{figure}

With the introduction of boundaries as  in the present case, we have an additional 
challenge. 
When there is a single phase error on any qubit other than the qubits on rough boundaries, exactly two checks are violated and two  nonzero syndromes created. When there is an error on a rough boundary qubit, then only one nonzero syndrome  is formed, 
see Fig.~\ref{fig:singleZerrors}. 
 An odd number of nonzero syndromes can be observed in the presence of  boundaries.
 
Note that the perfect matching algorithm requires an even number of nonzero syndromes, so it cannot be used
directly. 
Even if there are even number of syndromes if the perfect matching algorithm were used without any modifications,
it cannot correct the errors on the boundary qubits.
We adapt the algorithm proposed in \cite{wang10} for 2D codes. We discuss this algorithm next. 

Errors on the 3D toric code can be identified with paths in the lattice.
We allow the paths to contain half edges i.e. qubits on the boundary. 
Three cases arise. They are illustrated in Fig.~\ref{fig:singleZerrors}.

\begin{compactenum}[i)]
\item A path that terminates on two non-boundary qubits. Such a path flips exactly two checks. 
These checks are also the end points of the path.
\item A path that terminates on one boundary qubit and a non-boundary qubit. 
Such a path flips exactly one check. 
The violated check is an end point of the path. 
\item A path the terminates in two boundary qubits. 
In this case the path does not flip any check. 
This corresponds to an  error with zero syndrome. 
\end{compactenum}

To apply the matching algorithm we construct an auxiliary graph $\mathfrak{K}$ whose vertex set is the set of vertices with nonzero syndrome. Between any pair of nonzero syndrome nodes we add an edge whose weight is
the shortest distance between the two nodes i.e. the number of edges in the shortest path between the nodes.
We add a boundary node for every vertex in $\mathfrak{K}$  with nonzero syndrome. The edge connecting the vertex to the corresponding boundary node
has the weight of the shortest path connecting the node to the boundary. 
See Fig.~\ref{fig:completeGraph} for illustration. 
This allows the minimum weight perfect matching algorithm to find a path that involves the boundary qubits. 
We also add edges of zero weight between the boundary qubits. This will account for the case when the 
error does not involve the boundary qubits.  
The perfect matching algorithm will find a matching among the boundary nodes. These edges can be ignored 
when forming the associated error estimate. 
The graph  $\mathfrak{K}$  will always contain a minimum weight perfect matching. 
The complete procedure is given in Algorithm~\ref{alg:perfectmatch}. 
Note that the matching algorithm is a polynomial time algorithm.

\begin{algorithm}[H]
\caption{ Decoding phase errors on solid code. \cite{wang10}}
\begin{algorithmic}[1]
\REQUIRE {Syndrome for a phase error on the solid code.}
\ENSURE {Error estimate.}
\STATE Let $s_v$ be the syndrome on vertex $v$
\STATE Construct a graph $\mathfrak{K}$ whose vertex set is the set of  vertices with $s_v\neq 0$.
\FOR {$v$ with $s_v\neq 0$}
\FOR { $u\neq v$ and $s_u\neq 0$}
\STATE Find $P_{uv}$ the shortest path from $u$ to $v$.
\STATE Let $d_{uv}$ be the number of edges in $P_{uv}$
\STATE Add an edge $(u,v)$ in $\mathfrak{K}$   with weight $d_{uv}$
\ENDFOR
\STATE Add a new vertex $v'$ to $\mathfrak{K}$.
\STATE Find the shortest path to the boundary from $v$.
\STATE Let $d_{vv'}$ be the number of edges in the shortest path.
\STATE Add an edge connecting $v$ and $v'$ in $\mathfrak{K}$ with weight $d_{vv'}$
\ENDFOR
\STATE Form a complete graph on all the boundary nodes with each edge weight zero.
\STATE Find the minimum weight perfect matching on $\mathfrak{K}$.
\STATE Return the error corresponding to the matching as the error estimate. Ignore edges among  boundary nodes.
\end{algorithmic}
\label{alg:perfectmatch}
\end{algorithm}

\subsection{Correction of bit flip  errors} \label{section X errors}

\begin{figure}[H]
\centering
\includegraphics[scale=0.8]{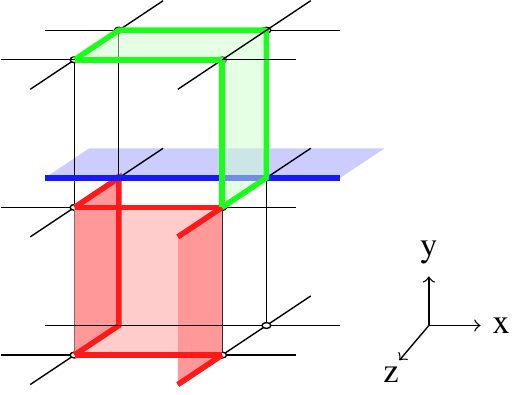}
\caption{Bit flip errors (in the dual lattice). Qubits in error are shown shaded. Nonzero syndromes are shown in solid color lines.}
\label{fig:connectederror}
\end{figure}

In this section we propose a local decoder for the bit flip errors. 
It is helpful to view the 3D toric code in dual lattice, see Fig.~\ref{fig:dualcode}. 
Now the qubits are on faces and the $Z$ type checks are on the edges. Errors correspond to faces and the syndrome is 
nonzero on the edges which form the boundary of the error. In standard 3D toric code with periodic boundary conditions,
nonzero syndromes will show up as cycles only. 
However, in the presence of boundaries the nonzero syndromes can show up as collection of cycles and paths. 
This is illustrated in Fig.~\ref{fig:connectederror}.
\begin{compactenum}[i)]
\item Errors only on the interior qubits i.e. those not on rough boundary. In this case non-zero syndromes form a closed boundary. 

\item If there is an error on the boundary qubits, then the syndrome is nonzero on an open string.
 Two such strings are shown in Fig.~\ref{fig:connectederror}.
\end{compactenum}

\begin{figure}[H]
        \begin{subfigure}[b]{0.4\textwidth}
                \centering
                \includegraphics[scale=0.5]{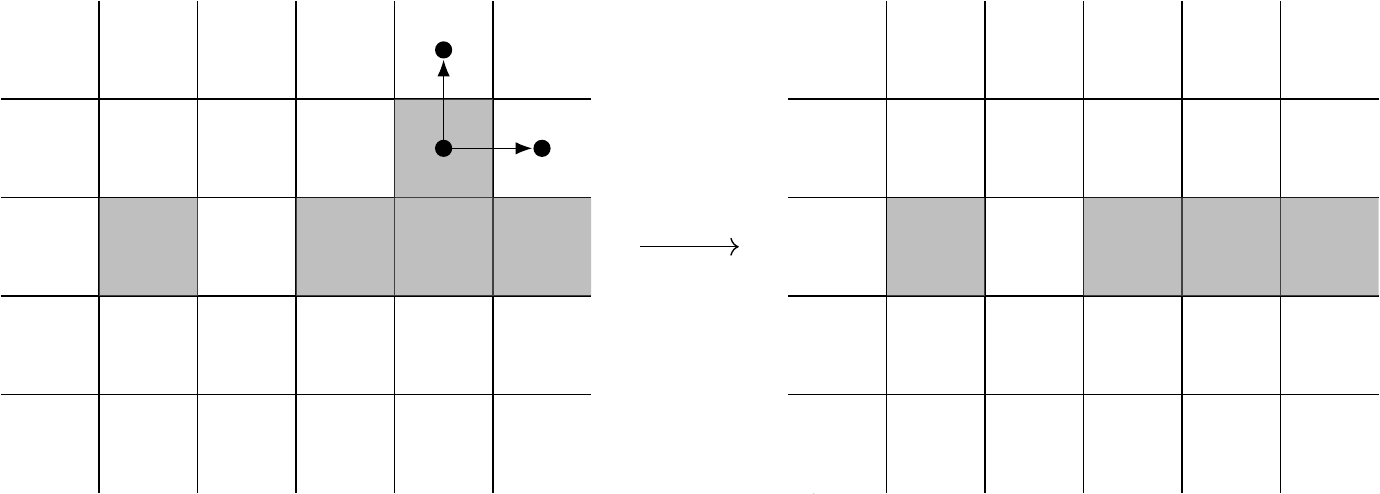}
				  \caption{}
				  \label{fig:ToomsRule1}
        \end{subfigure}%
        \vspace{0.5cm}
        \begin{subfigure}[b]{0.4\textwidth}
               \centering
				\includegraphics[scale=0.5]{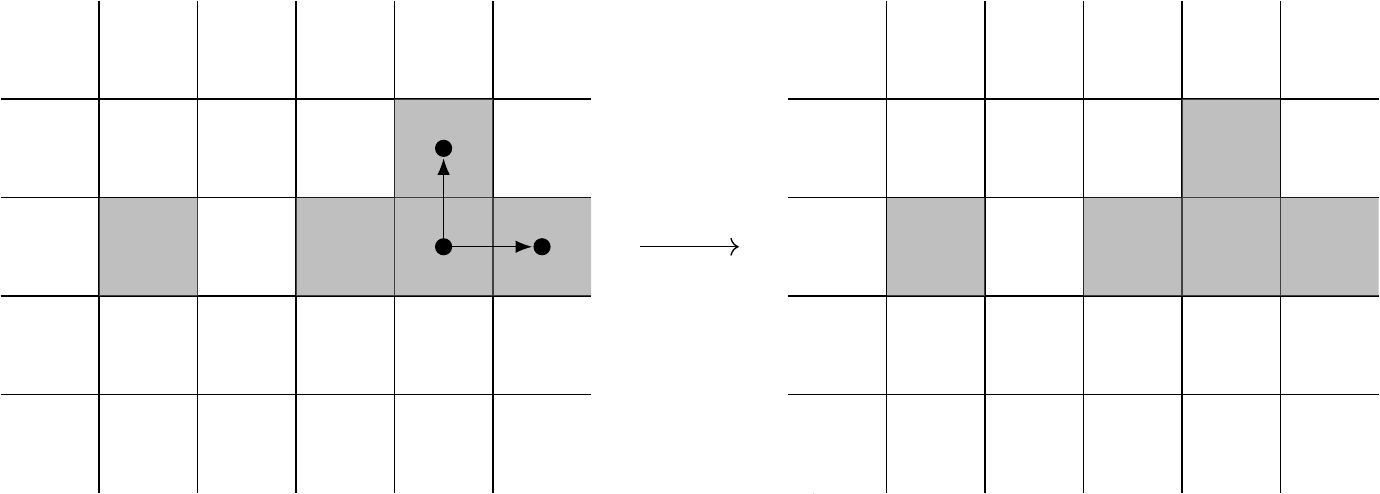}
				\caption{}
				\label{fig:ToomsRule2}
        \end{subfigure}%
        \vspace{0.5cm}
        \begin{subfigure}[b]{0.4\textwidth}
               \centering
				\includegraphics[scale=0.5]{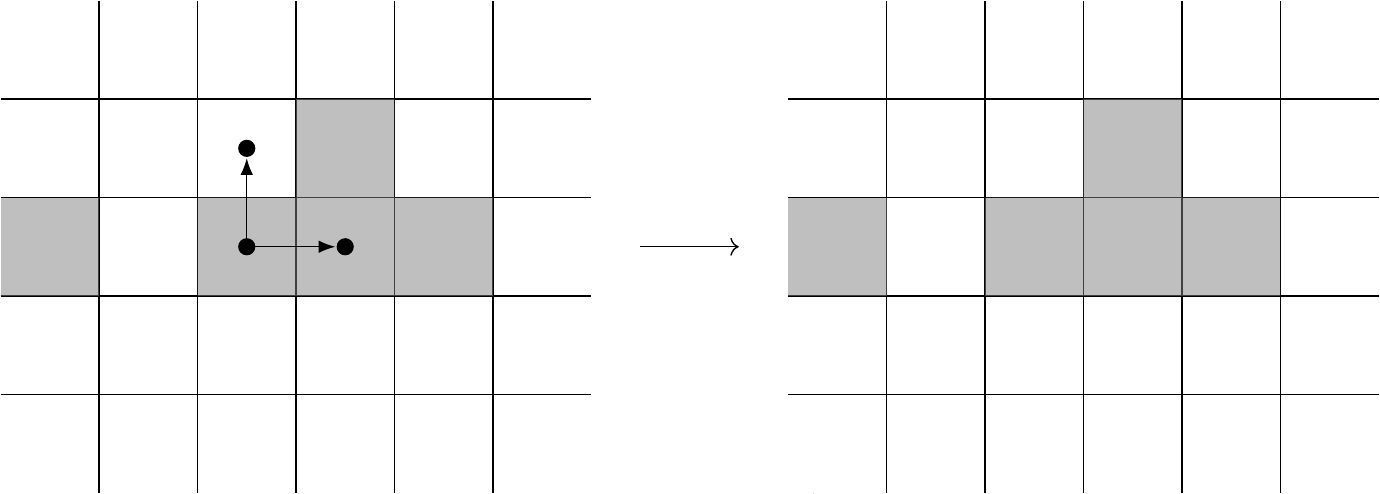}
				\caption{}
				\label{fig:ToomsRule3}
        \end{subfigure}
        \caption{Illustration of Toom's rule}\label{fig:toomsrule}
\end{figure}

The decoder for 3D code is motivated by the Toom's rule for classical 2D memories. 
The classical memory consists of a (periodic) square lattice with bits on every face. 
As per Toom's rule, a  cell is flipped if the neighboring cells on the north and east  have a different value. 
Thus,   the decoder takes the majority value of the bits in these three cells. 
We show application of Toom's rule by an example in  Fig.~\ref{fig:toomsrule}. 
The rule is applied to each cell from right to left and top to bottom. 
Fig.~\ref{fig:toomsrule} shows the configuration after the application of the rule on the marked cell. 

This rule has been adapted for qubits in the context of the 4D toric code in \cite{dennis02,pastawski11}.
In the quantum version we look at the syndrome on the north and east boundaries and flip the qubit if they are both 
nonzero. Ref.~\cite{dennis02} also made the rule probabilistic. 

\begin{figure}[H]
        \begin{subfigure}[b]{0.15\textwidth}
                \centering
                \includegraphics[scale=0.6]{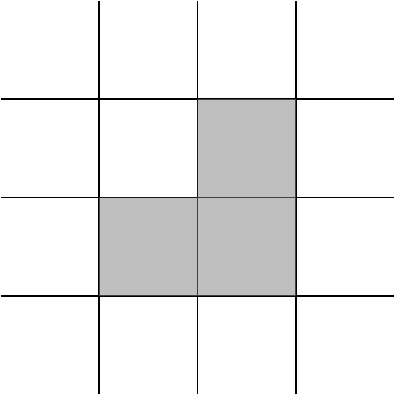}
				  \caption{}
				  \label{fig:QuantumToomsRule1}
        \end{subfigure}%
        \begin{subfigure}[b]{0.15\textwidth}
               \centering
				\includegraphics[scale=0.6]{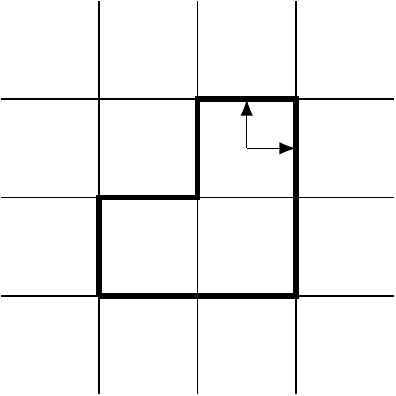}
				\caption{}
				\label{fig:QuantumToomsRule2}
        \end{subfigure}%
        \begin{subfigure}[b]{0.17\textwidth}
               \centering
				\includegraphics[scale=0.6]{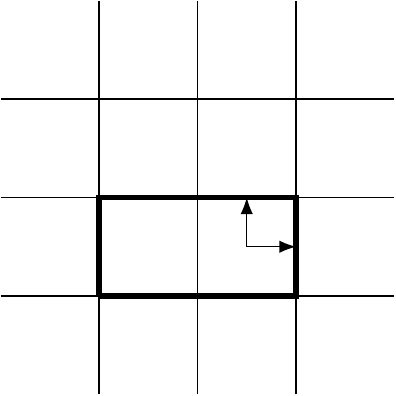}
				\caption{}
				\label{fig:QuantumToomsRule3}
        \end{subfigure}%
        \caption{Illustration of quantum version of Toom's rule. (a)  Initial error pattern (b)  Non-zero syndromes (in bold) (c) Non-zero syndromes pattern after applying rule to one cell.}\label{fig:QuantumToomsRule}
\end{figure}

We illustrate the quantum version of Toom's rule to an error pattern on $xz$ plane in solid code by an example in Fig.~\ref{fig:QuantumToomsRule}. For this error pattern we get non-zero syndromes on the boundary of the error as shown in Fig.~\ref{fig:QuantumToomsRule2}. We apply the rule sequentially to all the cells in the lattice.
At each cell we apply the the north-east rule. In Fig. \ref{fig:QuantumToomsRule3}, we show change in non-zero syndrome pattern after application of rule on one cell.

\begin{figure}[H]
        \begin{subfigure}[b]{0.17\textwidth}
                \centering
                \includegraphics[scale=0.6]{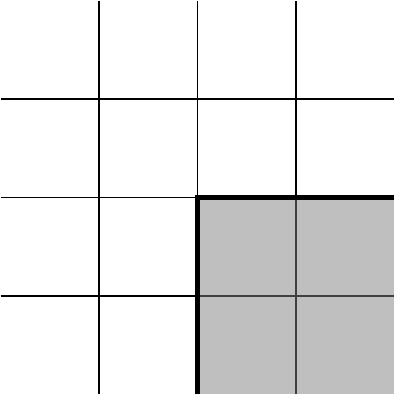}
				  \caption{}
				  \label{fig:NorthEastRuleFailure1}
        \end{subfigure}%
        \begin{subfigure}[b]{0.17\textwidth}
               \centering
				\includegraphics[scale=0.6]{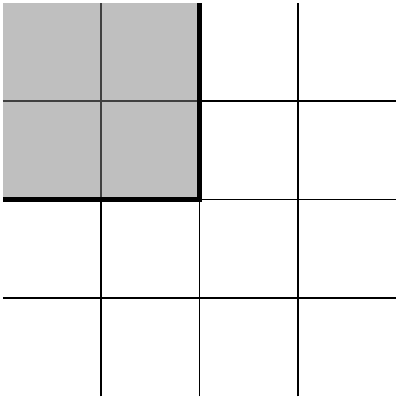}
				\caption{}
				\label{fig:NorthEastRuleFailure2}
        \end{subfigure}%
        \begin{subfigure}[b]{0.17\textwidth}
               \centering
				\includegraphics[scale=0.6]{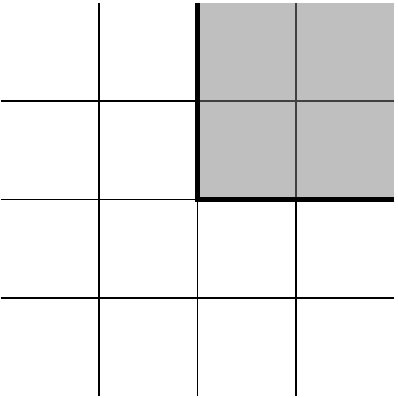}
				\caption{}
				\label{fig:NorthEastRuleFailure3}
        \end{subfigure}%
        \caption{Some error patterns where the  quantum version of Toom's rule fails. (The rule looks at the north and east boundaries.)}\label{fig:NorthEastRuleFailure}
\end{figure}

Ref.~\cite{breuckmann16} showed that there are certain error patterns in 4D toric code which cannot be corrected using the algorithm in \cite{dennis02}.  These patterns are persistent inspite of repeated application of the Toom's rule. 
A few such patterns are illustrated in Fig.~\ref{fig:NorthEastRuleFailure}. 

Error patterns shown in Fig. \ref{fig:NorthEastRuleFailure} are invariant under Toom's rule.
In order to correct these error patterns we introduce multiple rules instead of just one. 
If
Toom's rule were modified to consider north and west boundaries, then 
we can see that error pattern in Fig. \ref{fig:NorthEastRuleFailure1}  
can be corrected. 
However, the error pattern in Fig.~\ref{fig:NorthEastRuleFailure2} cannot be corrected by this modification. We need to consider yet another rule which looks at the south and east boundaries.

The failure of a single rule is overcome by considering alternate pair of boundaries of the cell. 
We are therefore naturally led to the idea of multiple local update rules. 
We propose to apply these rules sequentially.
More precisely, this means we first iterate with a particular local rule and see if the error pattern is corrected. 
If it is corrected, then we stop the decoder, otherwise we change the rule and run the decoder again. 
We repeat this process until all the rules are exhausted. 
Since a face in the 3D toric code can have at most four edges in its boundary, we can choose six pairs of edges to base the
Toom's rule. 

\begin{figure}[htb]
		\begin{subfigure}[b]{0.175\textwidth}
                \centering
               	\includegraphics[scale=0.8]{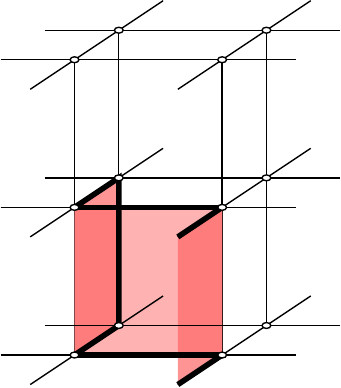}
				\caption{}
				\label{fig:step1}
        \end{subfigure}
		\begin{subfigure}[b]{0.17\textwidth}
                \centering
               	\includegraphics[scale=0.8]{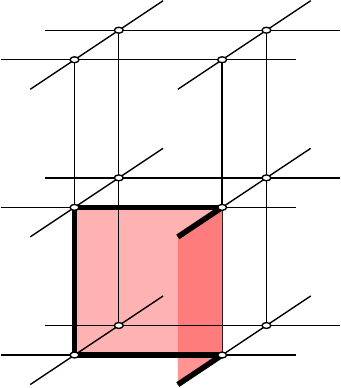}
				\caption{}
				\label{fig:step2}
        \end{subfigure}
		\begin{subfigure}[b]{0.17\textwidth}
                \centering
               	\includegraphics[scale=0.8]{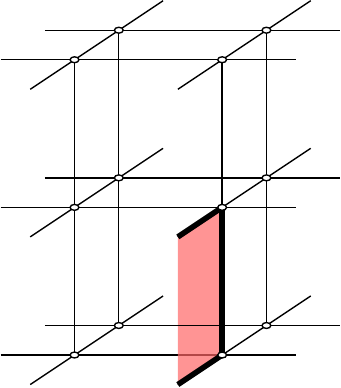}
				\caption{}
				\label{fig:step3}
        \end{subfigure}
		\begin{subfigure}[b]{0.17\textwidth}
                \centering
               	\includegraphics[scale=0.8]{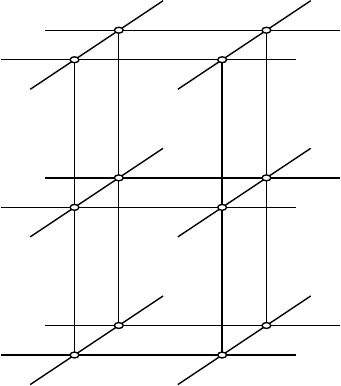}
				\caption{}
				\label{fig:step4}
        \end{subfigure}
        \captionsetup{justification=raggedright,singlelinecheck=false}
\caption{Illustration of bit flip error decoder.   Axes are oriented as in Fig.~\ref{fig:connectederror}. (a) Initial error pattern with non-zero syndromes.  (b) After updating  qubits parallel to $yz$ plane. (c)  After updating  qubits parallel to $xy$ plane.  (d) After updating  qubits parallel to $yz$ plane.}
\label{fig:exampleXerrors}
\end{figure}

Label the edges as $\mathsf{n, e, s, w}$ for the edges on the north, east, south and west. Then for a pair of edges 
$\alpha \beta\in \{\mathsf{ne, es, sw, wn, ns, ew} \}$, we apply $Z$ error to the  qubit if $\alpha $ and $ \beta$ edges have a  nonzero syndrome. 
For the boundary qubits without four edges, we ignore the rules involving the missing edges. 
We repeat this process for a fixed rule (i.e.  fixed $\alpha \beta$) for all the qubits according to some fixed sequence $\sigma$. 
For instance, we can go over all the planes parallel to $xy$-plane followed by planes parallel to $yz$ and $zx$-planes, and  in each plane left to right and top to bottom. 
Fig.~\ref{fig:exampleXerrors} illustrates how multiple rules can be applied to correct an error.

Unfortunately, there are error patterns which cannot be corrected even with multiple rules. 
One such error pattern is shown (in blue) in Fig.~\ref{fig:connectederror}.
Such errors need to be addressed separately.
These errors are characterized by nonzero syndrome which is not updated with the application of the rules. 
This happens when of the two edges being considered by the rule at most one edge has a nonzero syndrome. 
If a qubit has just one nonzero syndrome in its boundary then Toom's rule does not update the error estimate. 
The nonzero syndrome in its boundary must be cleared by the application of Toom's rule on its neighboring qubits. 
We can see that the error pattern in Fig.~\ref{fig:connectederror} will not be corrected  for this reason.

An error that cannot be corrected by all these local rules has a nonzero syndrome that is collection of strings. 
Each of these strings is parallel to either $x$ or $z$ axes or topologically equivalent to them. 
Oner such error is shown in blue in Fig.~\ref{fig:connectederror}.

Each such string partitions the  $xz$-plane containing the string into two sets. Flip all the qubits in the smaller set.
(This is not optimal, improvements are possible.) 
The complete listing is given in Algorithm~\ref{alg:Xdecoder}.

\begin{algorithm}[H]
\caption{{\ensuremath{\mbox{ Decoding  $X$ errors on solid code}}}}
\begin{algorithmic}[1]
\REQUIRE {Syndrome  $s$, for an $X$ error on the solid code, maximum number of iterations $I_{\max}$ and $J_{\max}$.}
\ENSURE {Error estimate, $E$.}
\STATE Let $E=I$
\WHILE { $s\neq0$ or $ i < I_{\max}$ } 
\FOR { $\alpha \beta\in \{\mathsf{ne, es, sw, wn, ns, ew} \}$}
\WHILE { $s\neq0$ or $s<J_{\max}$ }
\FOR {each qubit $q$ in a fixed sequence $\sigma$ }
\STATE Flip the qubit if there is nonzero syndrome
\STATE on the edges specified by $\alpha \beta$.
\STATE Update error estimate $E=E X_q$
\STATE Update  syndrome on the  edges of $q$. 
\ENDFOR
\ENDWHILE
\ENDFOR
\ENDWHILE
\IF  {$s\neq$ 0}
\FOR {each string $\kappa$ parallel to $x$ or $z$ axis}
\STATE   $\Omega =\{  \text{qubits in the $xz$ plane containing }\kappa \}$
\STATE  $\Omega_\kappa= \{\text{qubits in $\Omega$ to the left of } \kappa  \}$
\STATE Flip the qubits in smaller of the sets $\Omega_\kappa, \Omega\setminus \Omega_\kappa$
\STATE Update the error estimate and syndrome.
\ENDFOR
\ENDIF
\end{algorithmic}
\label{alg:Xdecoder}
\end{algorithm}

We empirically observed that for the decoder to clear all non-zero syndromes, value of $I_{\max}=\frac{\ell}{2}$ and that of $J_{\max}= \ell$. Decoder for the suggested values of $I_{max}$ and $J_{max}$ we observed that decoder clears all the syndromes. And increasing the $I_{max}$ and $J_{max}$ will not further improve the performance.
This is shown in Fig.~\ref{fig:CA_Study_of_L16} for $\ell=16$ linear length solid code. 
Heuristically this argument leads to complexity of the decoder to be $O(\ell^2n(\ell)) $ which is $O(\ell^5)$. 
Since $n(\ell)$ is $O(\ell^3)$, the complexity of the decoding algorithm is $O(n^{5/3})$.
The time complexity of the   algorithm can be reduced by parallelizing in lines 5--10.

\subsection{Simulation results}\label{sec:results}
\begin{figure}[H]
\centering
\includegraphics[scale=0.175]{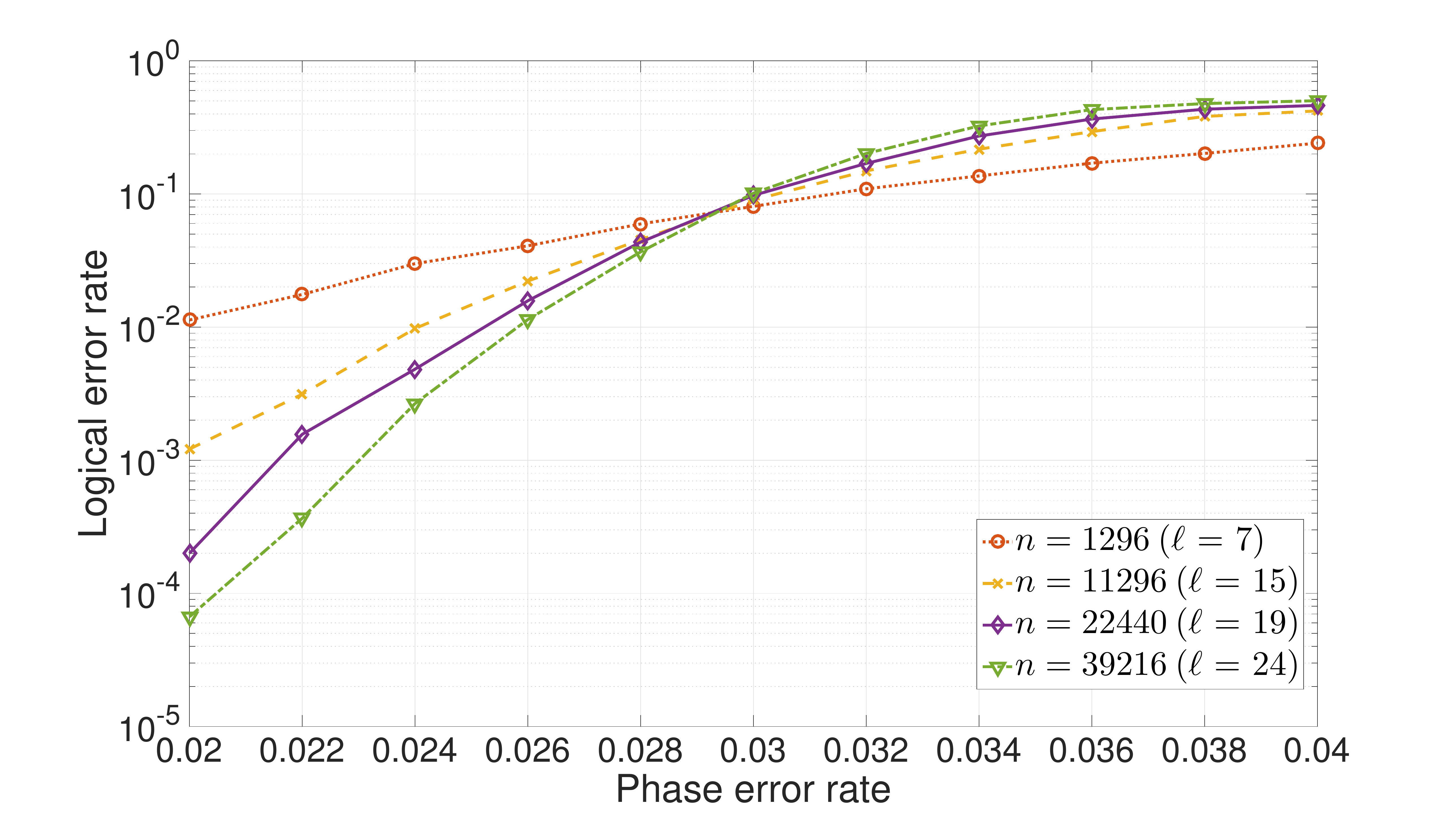}
\caption{Performance of Algorithm~\ref{alg:perfectmatch}  for phase  errors  on the 3D toric code with boundaries.} 
\label{fig:perfectmatchsimulation}
\end{figure}

\begin{figure}[H]
\centering
\includegraphics[scale=0.175]{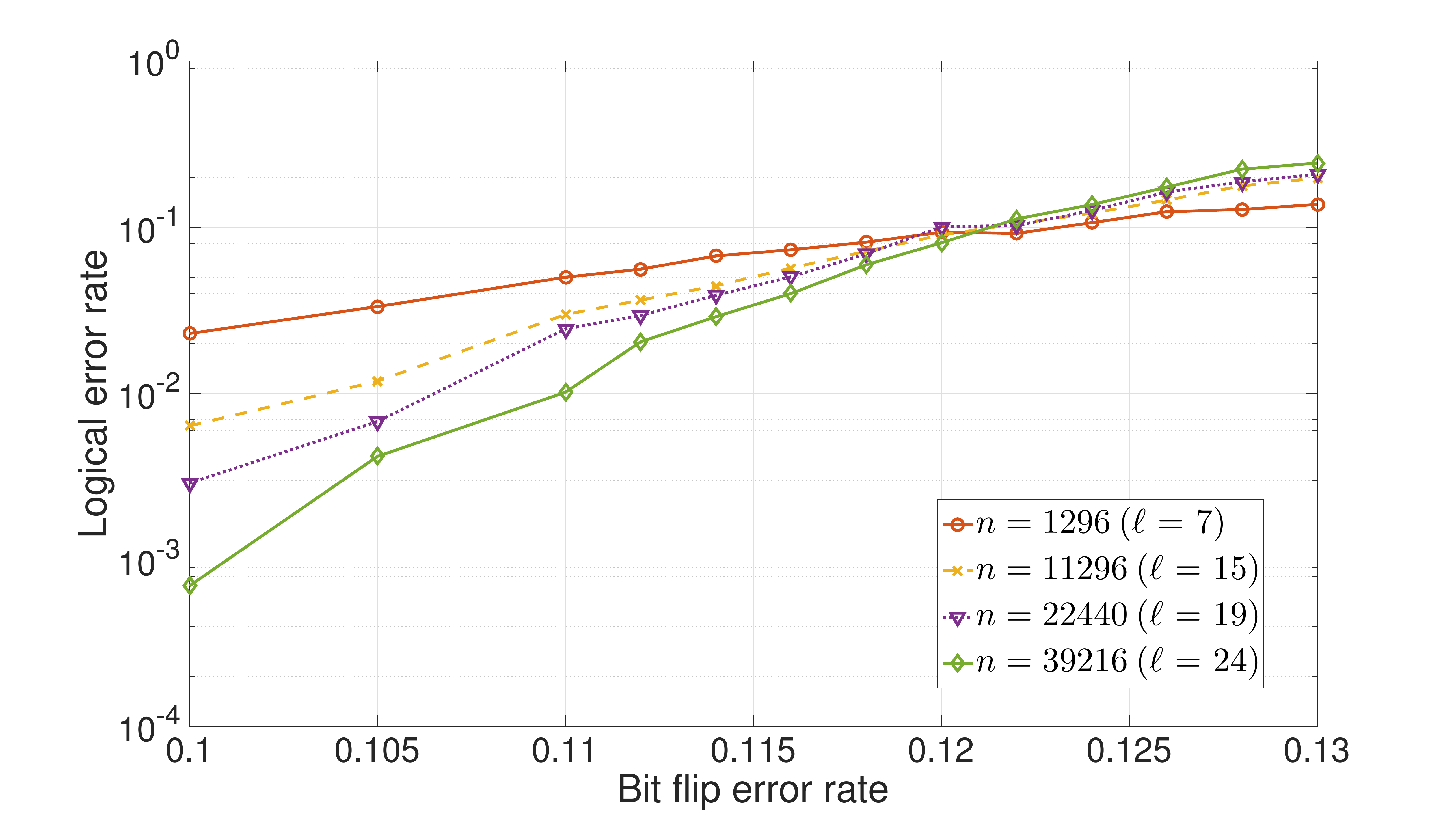}
\caption{Performance of Algorithm~\ref{alg:Xdecoder} for bit flip errors on the 3D toric code with boundaries.}
\label{fig:cellularsimulation}
\end{figure}

\begin{figure}[H]
\centering
\includegraphics[scale=0.175]{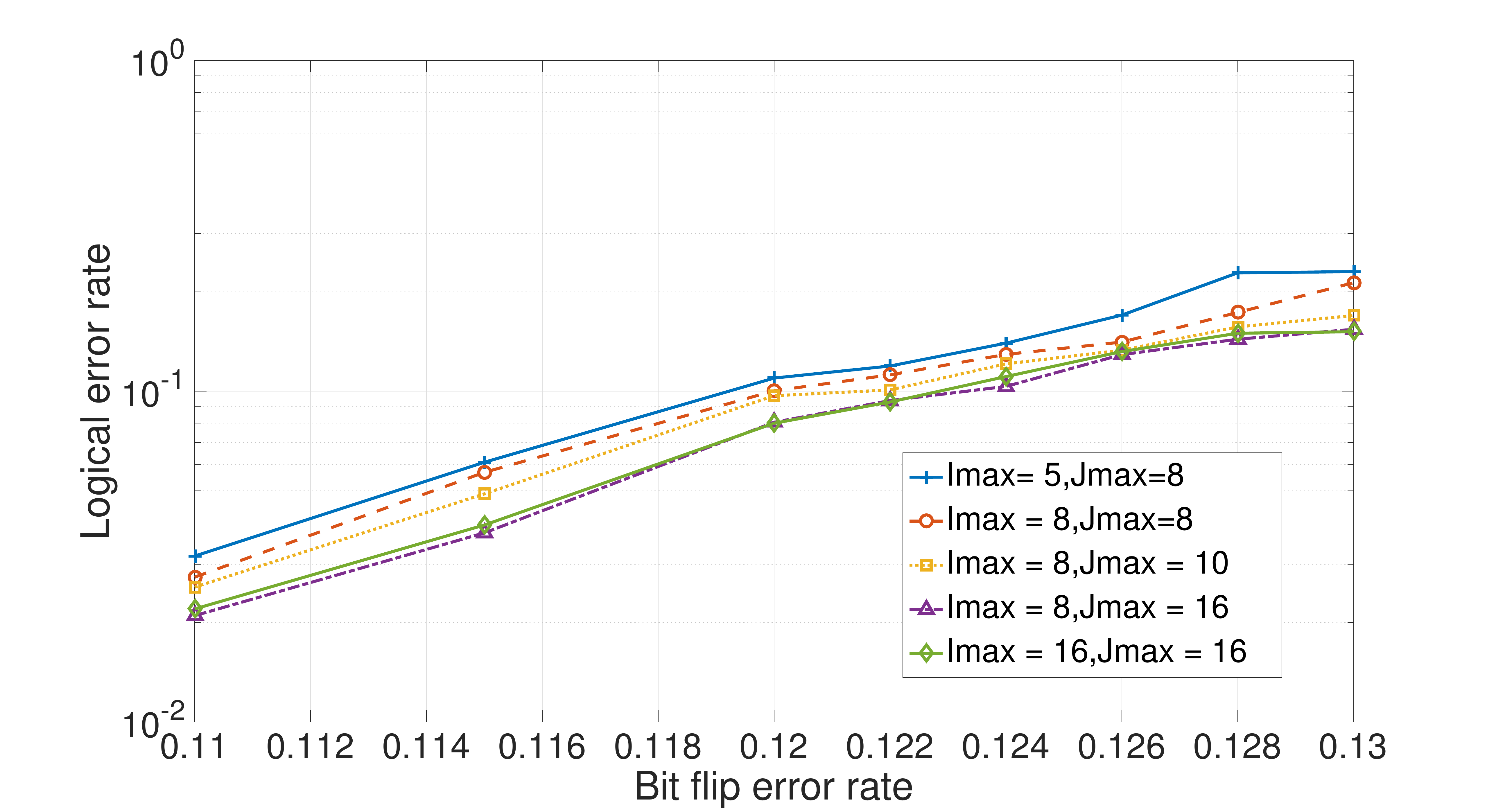}
\caption{Effect of 
$I_{\max}$ and $J_{\max}$ for solid code of size $\ell=16$ for various values of $I_{max}$ and $J_{max}$. We observed that decoder shows little or no improvement in performance beyond $I_{max}=\ell/2$ and $J_{max}=\ell$.} 
\label{fig:CA_Study_of_L16}
\end{figure}

In this section we report the performance of the decoders for 3D toric code with boundary. The matching decoder on the phase flip channel gives a threshold  $\gtrsim 2.9\%$, see Fig.~\ref{fig:perfectmatchsimulation}. 
  Algorithm~\ref{alg:Xdecoder} gives a threshold  $\gtrsim 12\%$ on the bit flip channel, see Fig.~\ref{fig:cellularsimulation}. 
  The effect of $I_{\max}$ and $J_{\max}$ are shown in  Fig.~\ref{fig:CA_Study_of_L16}.
Recall from Algorithm~\ref{alg:Xdecoder} that $J_{\max}$ is the number of times a given rule is applied while $I_{\max}$ is the number of times one round of application of all the rules each $J_{\max}$ times. 
After the completion of this work we came to know of the result by Duivenvoorden \etal \cite{duivenvoorden17} who proposed a renormalization decoder which gives a threshold of 17.2\% for the bit flip channel.
Ohno \etal estimate the threshold for phase errors to be 3.3\% \cite{ohno04}.
Takeda \etal conjecture the thresholds of the 3D toric code for the phase and bit flip errors  to be 3.46\% and 23.27\%  respectively \cite{takeda05}. 
  
 As we increase the number of times we iterate the rules 
 performance increases until $J_{\max}=\ell$.
 It appears that the number of times we need to cycle through all the rules is $\ell/2$.


\section{Decoding 3D toric code over the quantum erasure channel}\label{sec:3dtc-qec}

In this section, we propose a decoder for the 3D toric code over the quantum erasure channel. 
First we reformulate the erasure decoding algorithm \cite{delfosse17} in  
linear algebraic terms and Tanner graphs. This will be useful for decoding erasures
on other classes of quantum codes. 
We then consider the case the 3D toric code on the cubic lattice with periodic boundary conditions. 
Then we discuss the modifications when there are boundaries.

\subsection{An iterative decoding algorithm for erasures on CSS codes}
Recall that in the erasure channel a qubit is erased with probability $p$ and left as it is with probability $1-p$.
Letting $\rho$ be the state of the qubit, this channel can be modeled as 
\begin{eqnarray}
\mathcal{E}(\rho) = (1-p)\rho + p \ket{\kappa}\bra{\kappa},
\end{eqnarray}
where $\ket{\kappa}$ is a state orthogonal to the computational state space.

We replace each erased qubit by a qubit in the state, $I/2$ and measure the stabilizer generators. 
This has the effect of projecting a  Pauli error on each erased qubit uniformly at random. 
We call these Pauli errors {\em erasure induced errors}.

Let $\mathcal{E}$ be the set of erased qubits, $E$ be the induced Pauli error on the erased qubits.
Denote by $S_{\mathcal{E}}$, the stabilizers with support entirely in $\mathcal{E}$.
The (erasure) decoding problem is to estimate an error consistent with the syndrome $s$ and whose support is entirely in $\mathcal{E}$.
More precisely, we need to estimate the coset $ES_{\mathcal{E}}$ which is most likely given the syndrome $s$.
Delfosse et al. showed the following result in \cite{delfosse17}.
\begin{proposition}[Delfosse et al. \cite{delfosse17}]{\label{MLRule}}
Given an erasure pattern $\mathcal{E}$, and a measured syndrome $\sigma$, any coset of a Pauli error $E$ i) with support in $\mathcal{E}$  and ii)  consistent with the measured syndrome is a most likely coset. 
\end{proposition}

We can also represent $E$  as an element in $\F_2^{2n}$.
We decompose the error as $E=(a|b)$ where $e$ is the $X$ component and $f$ the $Z$ component of $E$. 
Let $S \in \F_2^{(n-k) \times 2n}$  be the stabilizer matrix  of the code; the code is assumed to be CSS.
\begin{eqnarray}
S= \left[ 
\begin{array}{cc}
H& 0 \\
0 &T
\end{array}
\right]
\end{eqnarray}
The syndrome of $E=(a|b) \in \F_2^{2n}$ is given by 
\begin{eqnarray}
H(b|a)^t = \left[ \begin{array}{c} Hb^t\\Ta^t \end{array}\right] = \left[ \begin{array}{c}\sigma\\ \tau \end{array}\right].
\end{eqnarray}
The syndrome for phase errors is given by $\sigma= Hb^t$ and 
the syndrome for bit flip errors by $\tau= Ta^t$.
Denote the restriction of $H$ to the qubits in $\mathcal{E}$ by $H_\mathcal{E}$.
Since the unerased qubits suffer no errors we have $a_{\bar{\mathcal{E}}}=b_{\bar{\mathcal{E}}}=0$.
This implies that 
$H_{\bar{\mathcal{E}}}b_{\bar{\mathcal{E}}}^t=0$ and
$T_{\bar{\mathcal{E}}}a_{\bar{\mathcal{E}}}^t=0$.
Therefore, the decoding problem reduces to solving the following system of equations: 
\begin{eqnarray}
H_{\mathcal{E}}b_{\mathcal{E}}^t = \sigma \label{eq:qec-3d-z-sys}\\
T_{\mathcal{E}}a_{\mathcal{E}}^t = \tau \label{eq:qec-3d-x-sys}
\end{eqnarray}
These systems of linear equations can be efficiently solved. 
We are more often interested in a decoder of linear time complexity. 
For this purpose it helps to look at these linear systems of equations closely. 
Since these system of equations arose in the context of an actual error, they are consistent and have at least one solution.
The following cases can arise.
 \begin{compactenum}[i)]
 \item $\mathcal{E}$ does not support any stabilizer or logical operator. In this case the error estimate is unique.
 \item $\mathcal{E}$ supports a stabilizer but does not support any logical operator. In this case any error estimate 
 consistent with the syndrome is equivalent to the actual error up to a stabilizer.
 \item $\mathcal{E}$ supports a logical operator.
 A stabilizer generator may or may not be supported. 
 In this case there is one or more logical operators in the support of $\mathcal{E}$.
 \end{compactenum}

By Proposition~\ref{MLRule}, all errors on the erased qubits are equally likely, hence there is at least 50\%
chance of a decoding error in the last case. One might as well ignore this case.  
We can do no better than randomly choosing any one of the possible errors consistent with the syndrome.
In the first and second  case any estimate that is consistent with the observed syndrome is correct estimate. 
In the first case the solution is unique while in the second case, the system of linear equations has multiple solutions.

So we shall focus on decoding correctly in the first two cases. 
Delfosse et al. \cite{delfosse17} solved this problem for the 2D surface codes.
Their algorithm is optimal and has linear time complexity.

For the 3D toric codes, this algorithm can be used for correcting the phase errors ie for solving the system of equations
corresponding to $H_{\mathcal{E}}b_{\mathcal{E}}^t = \sigma$.
but not for the $X$-errors ie the system of equations corresponding to $T_{\mathcal{E}}a_{\mathcal{E}}^t = \tau$.

Clearly, these equations can be efficiently solved using Gaussian elimination. 
However, we seek a more efficient algorithm. 
To this extent we shall exploit the structure of the equations a little more. 
It will hopefully, give a slightly different perspective on the results of \cite{delfosse16}.

First, notice that in a system $Ax=y$ if any of the equations contain only one variable, those equations  
can be solved very easily. 
The variables in those equations can then substituted in the remaining equations  to obtain a reduced system of equations. 
We can repeat this process until there are no more  equations with exactly one variable. 
At this point every equation contains two or more variables. 
If the system has a nonzero kernel, then we are able to set some subset of variables to arbitrary values and solve for the rest.

Assume that every variable occurs in two or more equations. 
Suppose that $Az=0$, then $x+z$ is also a solution to  $Ax=y$. 
Thus  for all $i$ in the support of $z$, there is a solution with $x_i=0$ or $x_i=1$.
Therefore, we can choose $x_i$ as a free variable and set it zero in the system of equations. 
This gives a smaller system of equations and if any single variable equations are created we solve for those variables
otherwise we find another variable in the support of the kernel and set it to zero and 
repeat this process.

The bottom line of this approach is that first we find a syndrome which is incident only on one erased qubit. 
In this case the measured syndrome is completely explained by the erased qubit incident on it. 
In linear algebraic terms,
we need to solve for an equation with exactly one variable. 
Once this variable is found, it is be updated 
in other equations where it appears. 
This process is called {\em peeling} and similar to the peeling decoding of classical low density parity check codes over the binary erasure channel. 

If we find that all check nodes are incident on two or more erased qubits, then we set the error on one of the 
erased qubits to identity. 
Such qubits must be in the support of a stabilizer (or logical operator) for estimates consistent with the syndrome. 
We call this process {\em freezing}. 
We say a qubit is {\em frozen} if the error estimate on it is (arbitrarily) set to identity.
In linear algebraic terms stabilizers and logical operators in with support in the erased qubits are elements of the kernel 
of the system of equations under consideration. 

We denote the parity check matrix restricted to the set of qubits in  $\mathcal{E}$ as $H_{\mathcal{E}}$.
We also denote the syndromes on these checks by 
$\syn(\mathcal{E})$.

\begin{algorithm}[H]
\caption{Peeling decoder for erasures}
\begin{algorithmic}[1]
\REQUIRE {Set of erasures $\mathcal{E}$,  Tanner graph $\mathcal{T_E}$ defined on $H_{\mathcal{E}}$,  and   $\syn(\mathcal{E})$.}
\ENSURE {
Error estimate for qubits in $\mathcal{E}'\subseteq \mathcal{E}$, unresolved erasure set $\mathcal{F} = \mathcal{E}\setminus \mathcal{E}'$
and  (updated)  $\syn(\mathcal{F})$.
}
\STATE $\mathcal{E}'=\emptyset$ and $\mathcal{T} = \mathcal{T}_\mathcal{E}$.
\WHILE { there is check $c$  of degree one} 

\STATE $\mathcal{E}'= \mathcal{E}'\cup \{q \}$ \COMMENT $q$ is the qubit connected to $c$
\STATE $e_q= s_c$ \COMMENT $e_q$ is the error on qubit $q$
\STATE Delete $c$ from  $\mathcal{T}$.

\STATE Update syndrome for all check nodes $v$ incident on $q$, ie set $s_v=s_v+x_q$. 
\STATE Delete $q$ and the edges incident on $q$ in $\mathcal{T}$
\ENDWHILE
\STATE Return $e_q$ for all $q\in \mathcal{E}'$, $\mathcal{T_{F}}=\mathcal{T}$
and $\syn(\mathcal{F})$.
\end{algorithmic}
\label{alg:peeling}
\end{algorithm}

\begin{remark}[Limitations of peeling]
In some cases it is possible that after peeling, the syndrome remains 
nonzero for some checks and no check is connected to exactly one erased qubit.
In this case we can either solve the system of equations which is likely to be much smaller than the original system of equations. 
\end{remark}

In case of toric code, it is possible to work on the original lattice on which the code is defined instead of the associated Tanner graph. 
The Tanner graph picture is useful when considering other classes of codes.

\begin{algorithm}[H]
\caption{Decoder for phase flip errors induced by erasures on CSS codes}
\begin{algorithmic}[1]
\REQUIRE { Stabilizer matrix $H_{\mathcal{E}}$,  erasure set $\mathcal{E}$, and syndromes $\sigma$. }
\ENSURE {Estimate $z$ consistent with measured syndrome $\sigma $ ie $H_{\mathcal{E}}z^t = \sigma $,}
\STATE $\mathcal{F} = \mathcal{E}$
\STATE Construct Tanner graph $\mathcal{T_F}$ based on $H_{\mathcal{F}}$

\STATE Find all independent stabilizers and logical operators in the support of $\mathcal{F}$
\STATE For each operator $o_i$, 
freeze a distinct qubit $q_i$ ie $z_{q_i}=0$
and update $\mathcal{F} = \mathcal{F}\setminus \{q_i\}$
\IF{nonzero 
syndromes exist}
\STATE Peel $\mathcal{T_F}$ using Algorithm~\ref{alg:peeling} \COMMENT Peeling updates the erasure set $\mathcal{F}$ and $\syn(\mathcal{F})$
\IF {$\syn(\mathcal{F})\neq0$ after peeling}
\STATE Solve for the system of equations 
$H_{\mathcal{F}} z_{\mathcal{F}}^t = \syn(\mathcal{F})$ 
\ENDIF
\ENDIF
\end{algorithmic}
\label{alg:css-z-decoder}
\end{algorithm}

\begin{remark}[Variations]
It is not necessary to perform peeling and freezing in separate steps. 
One could perform in an alternating fashion, freezing only when it is not possible to peel.
One variation is shown in Algorithm~\ref{alg:dec-z-alternating}
\end{remark}

\begin{algorithm}[H]
\caption{ Decoder for phase flip errors induced by erasures on CSS codes}
\begin{algorithmic}[1]
\REQUIRE { Stabilizer matrix $H_{\mathcal{E}}$,  erasure set $\mathcal{E}$, and syndromes $\sigma$. }
\ENSURE {Estimate $z$ consistent with measured syndrome $\sigma $ ie $H_{\mathcal{E}}z^t = \sigma $,}
\STATE Initialize $\mathcal{F}= \mathcal{E}$
\STATE Construct Tanner graph $\mathcal{T_F}$ based on $H_\mathcal{E}$
\WHILE{nonzero 
syndromes exist}
\STATE Peel $\mathcal{T_F}$ using Algorithm~\ref{alg:peeling} 
\STATE Find $\text{stab}(\mathcal{F})$ a stabilizer or logical operator within the support of current erased qubits $\mathcal{F}$
\IF {$\text{stab}(\mathcal{F}) \neq I$} 
\STATE In Tanner graph, $\mathcal{T_F}$, randomly freeze one qubit $q$ from the support of  $\text{stab}(\mathcal{F})$ obtained in line 4.
ie  set $z_q=0$
\STATE $\mathcal{F} = \mathcal{F}\setminus \{q\}$
\ELSE 
\STATE Solve for the system of equations on 
$H_{\mathcal{F}} z^t = \syn(\mathcal{F})$
\STATE Set $\mathcal{F} = \emptyset$
\ENDIF
\ENDWHILE
\end{algorithmic}
\label{alg:dec-z-alternating}
\end{algorithm}

\begin{remark}[Decoding $X$ errors induced by erasures]
For correcting $X$ errors, Algorithm~\ref{alg:css-z-decoder}~or~\ref{alg:dec-z-alternating} can be used  but with the input, 
$T_{\mathcal{E}}$ instead of $H_{\mathcal{E}}$, and $X$-syndromes $  \tau$ instead of $\sigma$. The algorithm returns estimate $x$ such that 
$T_{\mathcal{E}}x^t = \tau$.
\end{remark}

\subsection{Decoding erasures on the 3D toric code}

The process of freezing and peeling has a simple graphical interpretation in case of the toric codes.
Furthermore, the process of freezing can be performed first.  
In  case of phase errors, $H_{\mathcal{E}}$ is exactly the vertex-edge incidence matrix of the erased edges and the checks incident on erasures. 
Further, the elements of the kernel $H_{\mathcal{E}}$
are precisely the cycles of the lattice formed by erasures. 
One qubit per cycle is frozen. 
In \cite{delfosse17}, this process amounts to finding a spanning forest of the erased lattice on which the toric code is defined.
The erased lattice is the sublattice consisting of erased qubits and the checks affected by the erased qubits. 
Finding the spanning forest amounts to deciding which variables are frozen. 
The leaf nodes of the forest correspond to the syndromes where peeling is to be performed. 
Finding a spanning forest of the erased lattice is equivalent to finding the spanning forest of the Tanner graph and then removing all degree one qubit nodes. 
This amounts to freezing these qubits.

\begin{figure}[H]
    \centering
    \includegraphics[scale=0.85]{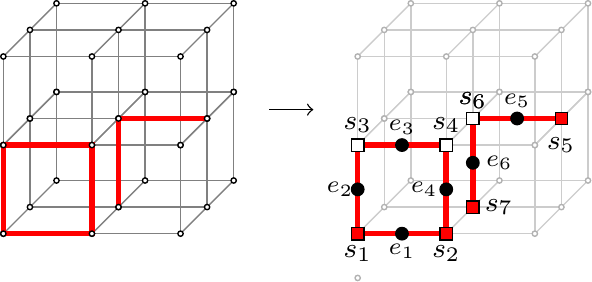}
    \caption{Erased qubits are shown in red.  The associated Tanner graph of the erased lattice is shown on the right where the qubits are shown by circles and syndromes by squares. Non-zero syndromes are shown by red squares.}
    \label{fig:TannerGraph}
\end{figure}

\begin{figure}[H]
    
    \includegraphics[scale=0.8]{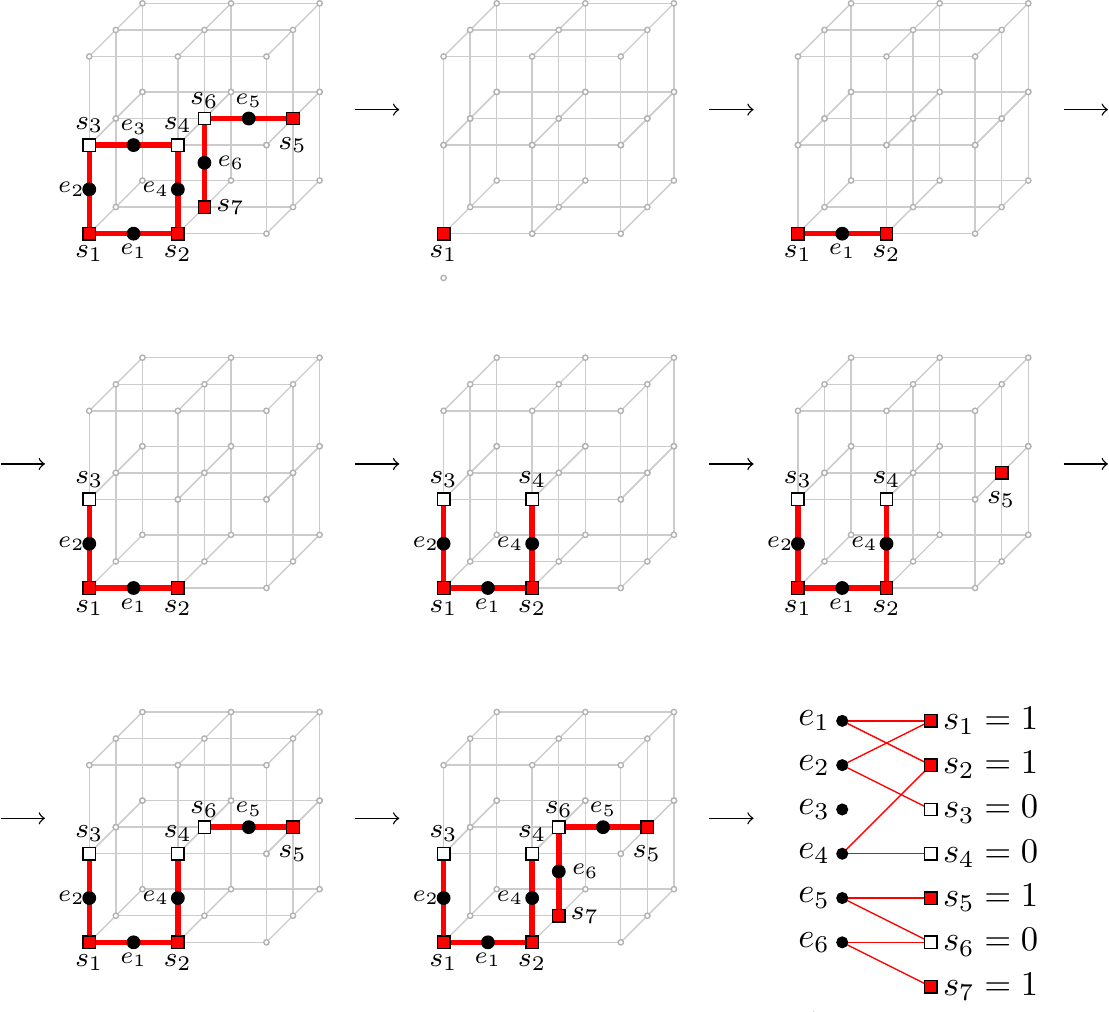}
    \caption{We show step by step construction of a spanning forest for the sublattice consisting of erased qubits $\mathcal{E}$.
    Equivalently, we can construct a spanning forest on the Tanner graph $\mathcal{T_E}$ and set the errors on qubits  with degree one  
    to be zero.}
    \label{fig:TannerGraph-forest}
\end{figure}

\begin{figure}[H]
    \centering
    \includegraphics[scale=0.85]{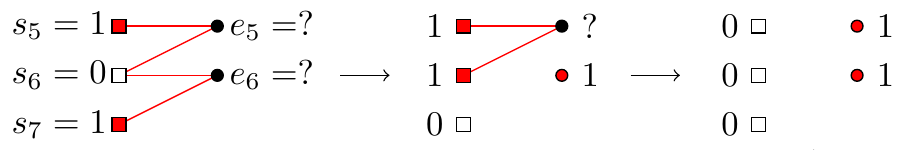}
    \caption{Illustrating the (partial) peeling of Tanner graph $\mathcal{T_E}$ after freezing as per Fig.~\ref{fig:TannerGraph}}
    \label{fig:peelingExample}
\end{figure}

The algorithm proposed in \cite{delfosse17} can be used to correct (erasure induced) phase errors for the 3D toric code. 
We illustrate this with the following example. Fig.~\ref{fig:TannerGraph} shows an erasure patttern and the associated Tanner graph. Fig.~\ref{fig:TannerGraph-forest} shows the freezing by contructing a forest of the Tanner graph. 
Next step in decoding is peeling. Peeling of Tanner graph 	after freezing is shown in Fig. \ref{fig:peelingExample}.


\begin{figure}[H]
    \centering
    \includegraphics[scale=1]{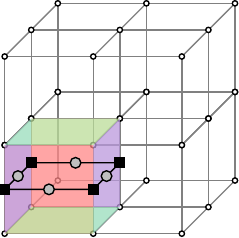}
    \caption{Erased qubits are shown in colored faces and they form an $X$-stabilizer. However, a cycle in the associated Tanner graph (shown partially) of this erasure pattern does not correspond to an $X$-type stabilizer. Squares represent the check node and circles represent qubit nodes of the Tanner graph.}
    \label{fig:cycle-nonstab}
\end{figure}

Finding the qubits to be frozen using the spanning forest approach does not work for correcting the bit flip errors in the 3D toric codes.
The difficulty is deciding which variables to freeze. 
While the 2D case allows us to simply find a cycle in the support of erased qubits and freeze any one of them,
that approach fails because cycles no longer correspond to (X-type) stabilizers in the 3D case. 
For instance, consider the erasure pattern shown in Fig.~\ref{fig:cycle-nonstab}.
All six faces of a unit cube are erased. 
In the Tanner graph associated to this pattern, there is a cycle which does
not correspond to a stabilizer. 
This cycle is shown in 
Fig.~\ref{fig:cycle-nonstab}. (Please note the entire Tanner graph is not shown).

We propose an algorithm for efficiently finding the qubits to be frozen. 
In case of $X$ errors, the elements of the kernel (of $T_{\mathcal{E}}$, see Eq.~\eqref{eq:qec-3d-x-sys}) are best visualized in the dual lattice. 
They are surfaces without boundaries in the dual lattice of the toric code. 
Stabilizers correspond to surfaces of trivial homology ie they are boundaries of closed volumes in the lattice.
Logical operators correspond to surfaces of nontrivial homology. 
To find these stabilizers, we can take the following approach. 

In the dual lattice $\Lambda^*$, delete all the qubits (faces) corresponding to erasures. 
This creates a collection of connected components in $\Lambda^*$ (and also in $\Lambda$).
Suppose we let a particle explore the lattice so that it can move from one cell to another only if they share an
unerased qubit. 
Let $\Omega_c$ be the collection of cells visited by particle starting from cell $c$.
If $\Omega_c$ is a closed volume then the boundary of the volume is precisely the stabilizer in that support of the erased qubits.
After finding a stabilizer, we start exploring the lattice from a cell that is not in $\Omega_c$ and
proceed to find other stabilizers until all the cells  are visited. 
We call this procedure  {\em the trapping algorithm}.

The trapping algorithm can be be also performed on the primal lattice, but the topological nature of errors is clearer in the dual lattice. 
In the primal lattice it is equivalent to finding a spanning forest in the 
unerased lattice i.e.  the 3D lattice obtained by deleting the erased qubits.

\begin{figure}[H]
\centering
\includegraphics[scale=0.75]{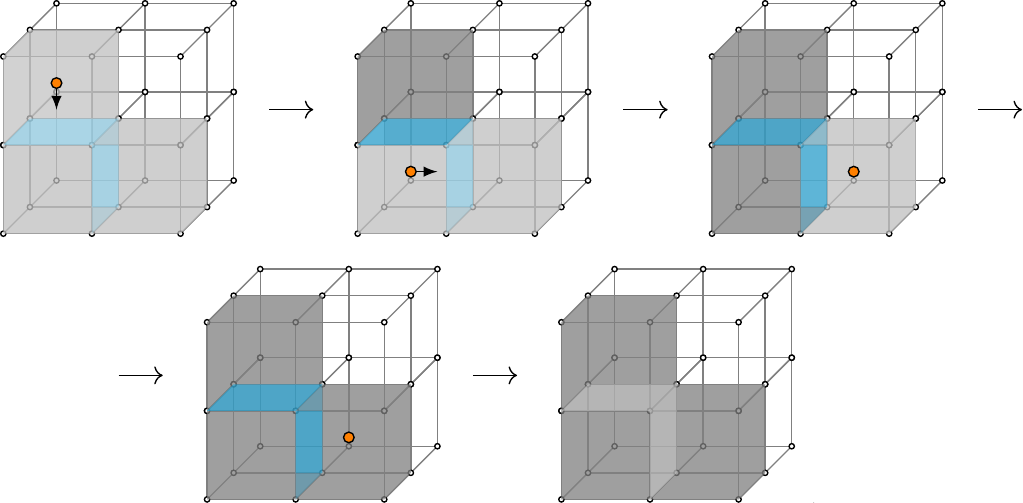}
\caption{An $X$-stabilizer corresponds to closed volume and boundary is the support of the stabilizer. 
We illustrate the algorithm to find stabilizer support.
Faces in light gray are the erased qubits. 
A particle can move from one cell to another by unerased faces. 
If the particles is trapped inside, then the boundary of that volume is the support of the stabilizer. Illustrated is the evolution of this process
for a simple volume. }
\label{fig:Trapping}
\end{figure}

To understand the behaviour of the decoder we need to consider the following types of erasure patterns. 
\begin{compactenum}[(a)]
\item 
If there is a stabilizer in the support of the erasure pattern, as in Fig.~\ref{fig:Trapping}, then the algorithm will recover the boundary of the volume corresponding to the
stabilizer. 
We can obtain all the independent stabilizers. 
Then the decoder can freeze a distinct  qubit in the boundary of each stabilizer and try to peel. Sometimes it may not be possible to peel after freezing.

\item 
If the erasure pattern contains the support of a logical $X$ operator as in Fig.~\ref{fig:TrappingLogical}, then 
the algorithm cannot recover its support. 
This is because the logical $X$ operators do not form a closed volume. 
For instance, in Fig. \ref{fig:TrappingLogical} if we start the start the algorithm from any unit cube, we will be able to visit all the remaining unit cubes. Therefore, X logical operator in support of erased qubits remain undetected. 

\begin{figure}[H]
\centering
\includegraphics[scale=1.3]{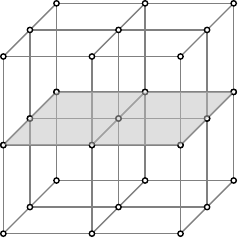}
\caption{The trapping algorithm when $X$ logical operator is in the support of erased qubits. It returns an empty boundary. Any one of the erased qubits can be frozen in this case.
There is at least 50\% probability of incorrect decoding.}
\label{fig:TrappingLogical}
\end{figure}

When the algorithm returns an empty boundary, we can freeze one of the erased qubits and try to peel. 
We repeat this process until we clear the syndrome. 
If we freeze correctly, then the decoder succeeds, if not, we have a logical error.
Both the outcomes are equally likely, 
if there is exactly one logical operator in support of erased qubits.
 So there is at most fifty percent chance to correct the error. In cases where there are more than one logical operator in support of erased qubit, number of cosets increase, thereby decreasing the probability of correcting the error, as all cosets are equally likely.
Considering the high probability of decoding incorrectly, the decoder might as well choose to declare a decoding failure and abort. 
This does not affect the performance substantially.
\item 
Another pattern where the algorithm returns an empty set is shown in Fig.~\ref{fig:KleinBottle}. 
All faces except ones marked in green are erased.
We call this pattern a pseudo Klein bottle pattern. 
This pattern does not contain the support of a stabilizer or logical operator. 
Again, because every edge (ie syndrome) participates in at least two qubits, peeling cannot be carried out. 
We have to correct $X$ errors in this erasure pattern exactly, correction up to a stabilizer is not possible. 
\begin{figure}[H]
\centering
\includegraphics[scale=0.8]{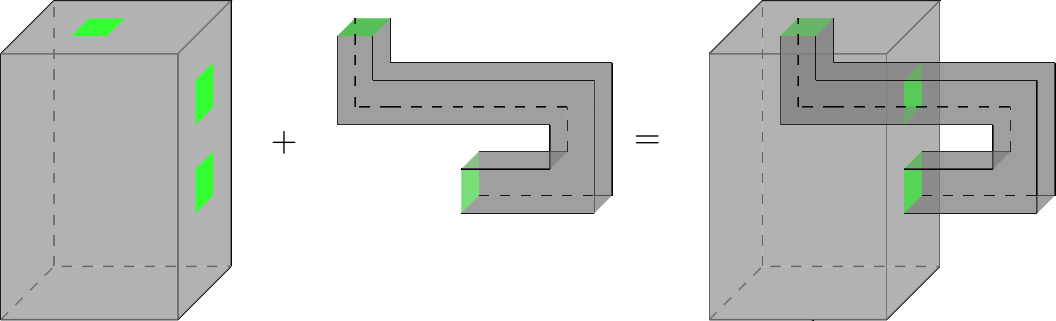}
\caption{Pseudo Klein bottle erasure pattern. All qubits on the boundary of the cuboid and the tube-like structure are erased except the ones in (green) color.  An erasure pattern like this returns an empty boundary from the trapping algorithm. There is a unique error which explains the observed syndrome. Freezing a qubit incorrectly leads to syndrome not being cleared by peeling.
}
\label{fig:KleinBottle}
\end{figure}

When the pseudo Klein bottle erasure pattern occurs, there is a unique error but the peeling procedure does not work, as every check is incident on at least two erased qubits.
There are at least three ways to proceed with the decoding as explained below. 
\begin{compactenum}[i)]

\item We could simply solve the system of linear equations corresponding to the residual erasure pattern at this juncture. 
\item We could randomly freeze a qubit and start peeling. If the qubit was frozen correctly, then we decode correctly. 
If we had frozen it incorrectly, the syndrome will not be cleared. At this point we could either backtrack or repeat the peeling. 
Alternatively, we could try to clear the syndrome by absorbing this error into a nonerased qubit. 
\item We could simply ignore such cases and declare a decoding failure. Our simulations shows that this does not limit the performance, since the decoder's performance is limited by the performance of the $Z$-type decoder. 
\end{compactenum}
In our implementation, whenever an erasure pattern contains one or more  pseudo Klein bottles  and logical operators, we consider it as a decoder failure. It is a maximum likelihood (ML) decoder in cases where decoder returns an estimate consistent with the syndrome. 
\end{compactenum}

\subsection{Erasure decoding of 3D toric code with boundaries}

We now show how to decode 3D toric code in the presence of boundaries, see Fig.~\ref{fig:solidcodeWithZ}.  
The decoders for phase error and bit flip error correction presented for 3D toric codes in previous section have to be modified to incorporate boundaries.

\begin{figure}[H]
\centering
\includegraphics[scale=0.7]{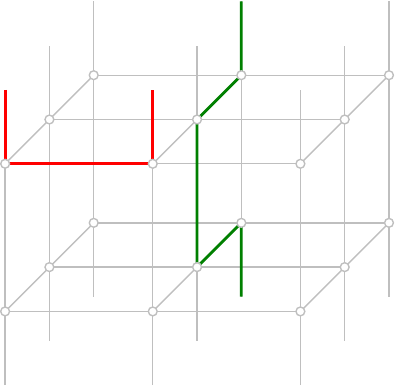}
\caption{Stabilizer generator (in red) and a logical operator (in green) with boundary qubits. These patterns cannot be peeled because there is no check node of degree one.}
\label{fig:3d-tc-stabilizers}
\end{figure}

First we describe how to incorporate boundaries for phase error correction.
In the presence of boundaries, some stabilizers no longer correspond to cycles in the lattice.
One such stabilizer is shown in Fig.~\ref{fig:3d-tc-stabilizers}. 
If we form the Tanner graph associated to this erasure pattern, peeling cannot proceed.
The spanning tree for the erasure pattern 
(in red) is itself. 
This pattern cannot be peeled, because there is no check node of degree one.
There is a stabilizer in the support of the erasure pattern but it is not a cycle and the algorithm discussed in the previous
section will fail in this case.
A similar problem exists when the erased qubits  corresponds to a logical operator. 
The peeling decoder gets stuck because there are no check nodes of degree one.

\begin{figure}[H]
		\begin{subfigure}[b]{0.22\textwidth}
                \centering
               	\includegraphics[scale=0.7]{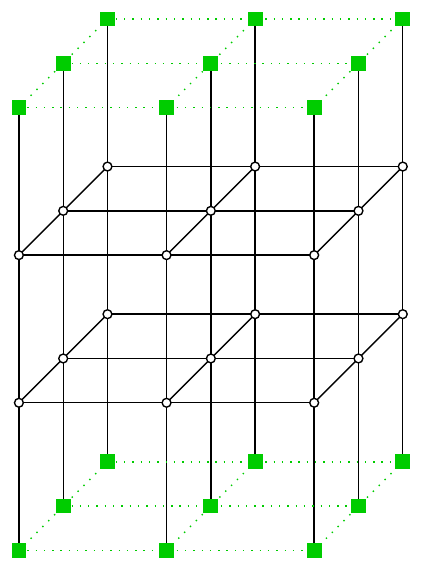}
				\caption{}
				\label{fig:SolidCodeComplete}
        \end{subfigure}
        \begin{subfigure}[b]{0.22\textwidth}
                \centering
               	\includegraphics[scale=1]{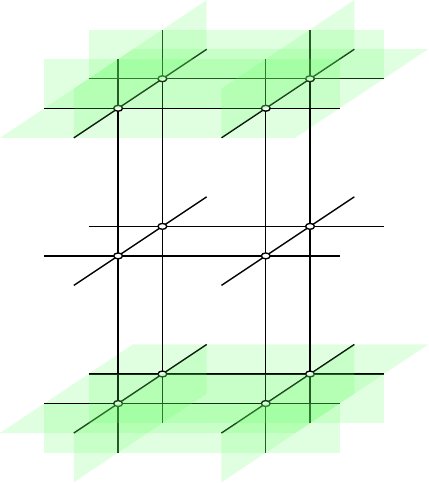}
				\caption{}
				\label{fig:SolidCodeCompleteDual}
        \end{subfigure}
\caption{(a) Solid code with dummy vertices (solid squares) and dummy qubits (dashed edges).  (b) Dummy vertices in the primal lattice (a) become dummy volumes and dummy edges become dummy faces (shown in green) in the dual lattice.}
\end{figure}

To resolve this problem in 2D toric codes Delfosse et al. \cite{delfosse17} add dummy vertices (checks) and edges (qubits) and to ensure  that peeling does not begin from a dummy check node require that a tree is grown  rooted at a dummy check node and does not contain any more dummy check nodes. 
This solution carries over to the 3D toric code for the phase errors. 

Similarly, in 3D toric code with boundaries we introduce dummy vertices and dummy edges, see Fig.~\ref{fig:SolidCodeComplete} for an illustration.
Dummy vertices carry no $X$-stabilizers and dummy edges carry no qubits on them.

\begin{figure}[H]
\centering
\includegraphics[scale=0.7]{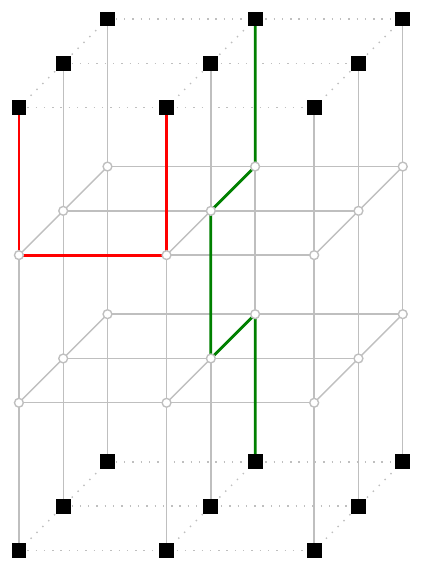}
\caption{A string with two dummy nodes is either a $Z$ stabilizer (shown in red) or a $Z$ logical operator (shown in green).}
\label{fig:solidCodeTrivialLoop}
\end{figure}

Syndromes never occur at dummy vertices. 
Spanning forest of the Tanner graph is constructed rooted at dummy check nodes and  with an additional condition that any connected component cannot have more than one dummy vertex, which means it cannot have more than one rough boundary qubit. This is because a string with two dummy nodes will either form $Z$ stabilizer or $Z$ logical operator as illustrated in Fig. \ref{fig:solidCodeTrivialLoop}. 

\begin{figure}[H]
		\begin{subfigure}[b]{0.22\textwidth}
                \centering
               	\includegraphics[scale=0.8]{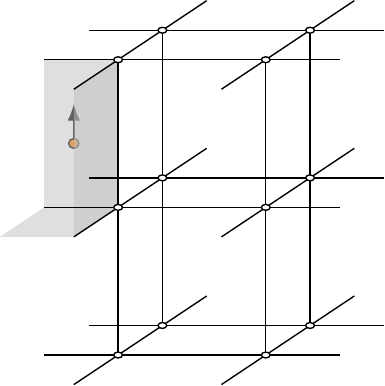}
				\caption{}
				\label{fig:trapping-failure-1}
        \end{subfigure}
        \begin{subfigure}[b]{0.22\textwidth}
                \centering
               	\includegraphics[scale=0.8]{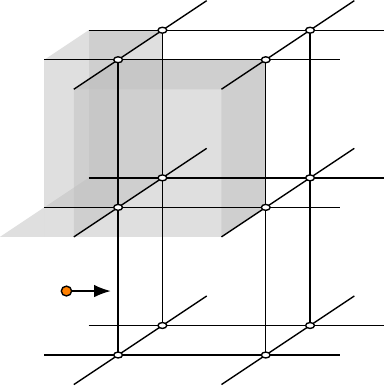}
				\caption{}
				\label{fig:trapping-failure-2}
        \end{subfigure}
\caption{Some represenative erasure patterns which cause trapping algorithm to fail. These are resolved by adding dummy qubits and checks. (a) An erasure pattern where non-stabilizer erasure pattern will be returned as a boundary by {trapping} algorithm. (b) An erasure pattern where partial boundary of the stabilizer plus a qubit not a part of stabilizer is returned as boundary by {trapping} algorithm. }\label{fig:trapping-failure}
\end{figure}

Next we describe how to incorporate boundaries for bit-flip error correction. 
The idea behind the  trapping algorithm is to
let a particle explore the lattice via unerased qubits and return the boundary of volume to which the particle  is confined.
Since there are boundaries, even if there are  no erasures, the particle is confined between the boundaries. 
So running the trapping algorithm on the (dual) lattice can cause
the algorithm to fail. 
Even if choose to ignore the unerased qubits confining the particle, 
there are also other problems due to boundaries which cause the  trapping algorithm to fail.
Fig.~\ref{fig:trapping-failure}
illustrates some representative cases. 

Fortunately, there is a simple solution. 
We only need to add dummy qubits (which are never erased) and dummy checks. The modified lattice and its dual with the dummy qubits and checks are shown in 
Fig.~\ref{fig:SolidCodeComplete} and Fig.~\ref{fig:SolidCodeCompleteDual}, respectively. Dummy vertices now form dummy volumes in dual and dummy edges form dummy faces. 
Then the trapping algorithm, as discussed for the periodic lattice can be used without any problems. 

\begin{figure}[H]
\centering
\includegraphics[scale=1]{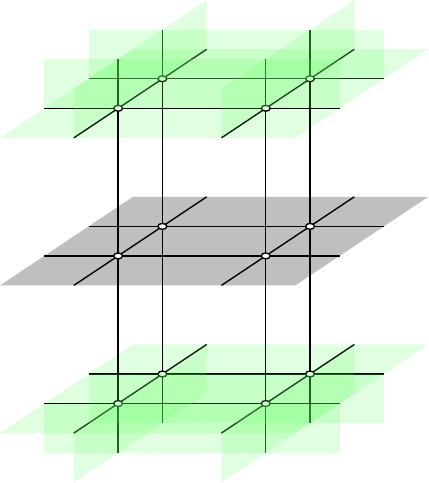}
\caption{Erased qubits are shown in gray color. Dummy qubits are shown in green. Since solid code has boundary a particle gets trapped to either  above or below the plane of erased qubits.}
\label{fig:TrappingLogicalSolidCode}
\end{figure}

Aperiodicity in the solid code gives an added advantage, compared to the toric code on the periodic lattice. 
We can determine if the erasure pattern supports an
$X$ logical operator
using trapping algorithm. (This is unlike the periodic boundary case.) 
We illustrate this with an example in Fig. \ref{fig:TrappingLogicalSolidCode}. Also shown is an $X$ logical operator in support of erased qubits.
The gray planes separate the actual code from the dummy volumes. Gray plane is the erasure pattern. In this case, we can  see that if we start the trapping algorithm from any unit cube which is  above the gray plane, we get trapped in volumes above the gray plane. And vice-versa if we start from above the plane. 
In case of the periodic lattice, the trapping algorithm returns an empty boundary.  However, with boundaries the trapping algorithm returns the support of the logical $X$ operator.

\subsection{Simulation results} 
\begin{figure}[H]
\includegraphics[scale=0.175]{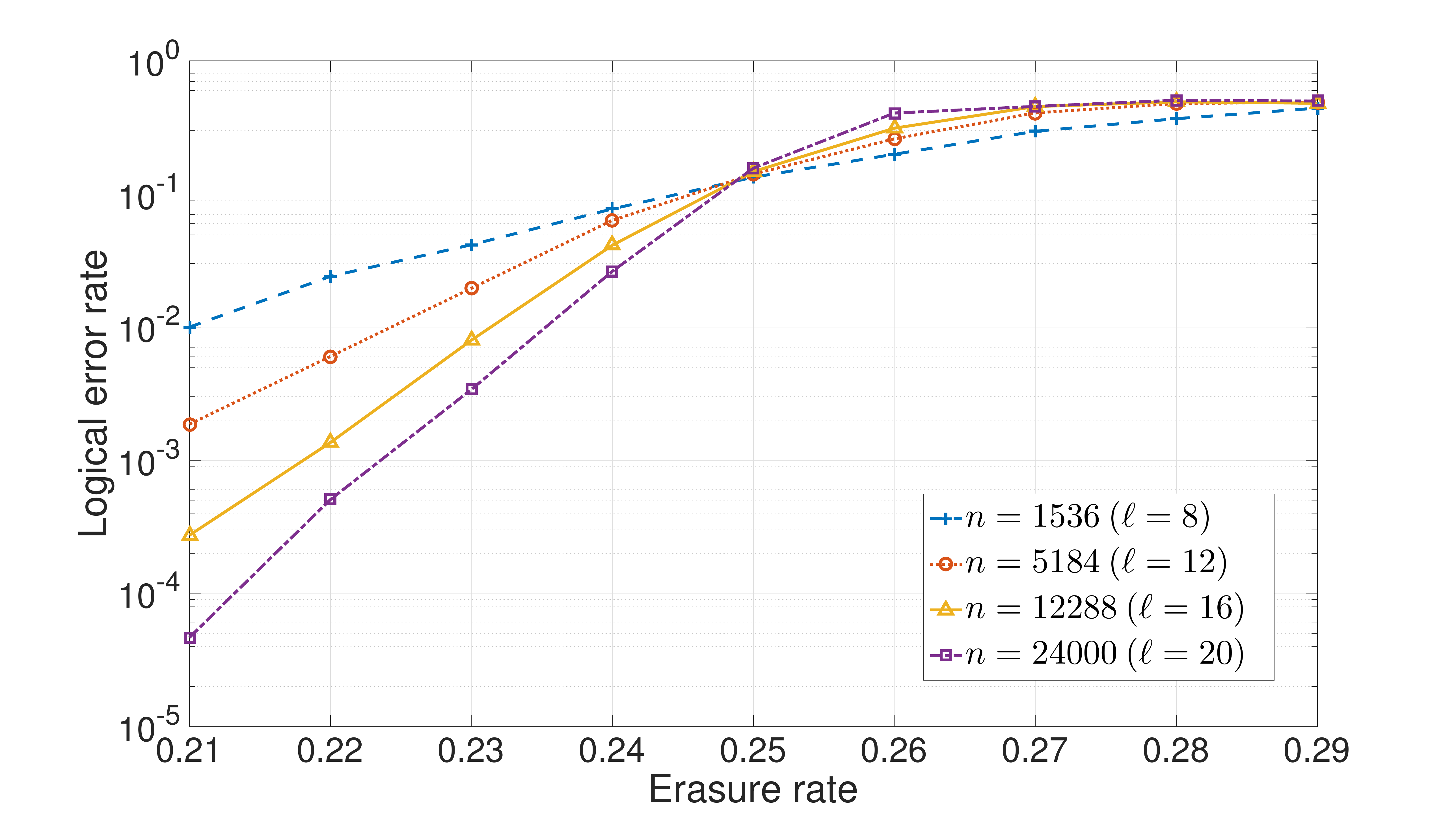}
\caption{Performance of the 3D toric code with periodic boundary over the quantum erasure channel.}
\label{fig:xz-sims-3dtc}
\end{figure}

In this section we present performance of 3D toric code with and without boundaries. 
Fig.~\ref{fig:xz-sims-3dtc} shows the performance of the 3D toric code with periodic boundary conditions.
This is essentially the same as the performance of the 3D toric code with respect to the erasure induced $Z$ errors. 
The overall erasure threshold is therefore about 24.8\%.
We note that this is quite close to the bond percolation threshold for the cubic lattice \cite{wilke83}. 
Similar observations have been made for the 2D toric codes \cite{dennis02} and codes over hyperbolic   tilings \cite{delfosse13}.

\begin{figure}[H]
\includegraphics[scale=0.175]{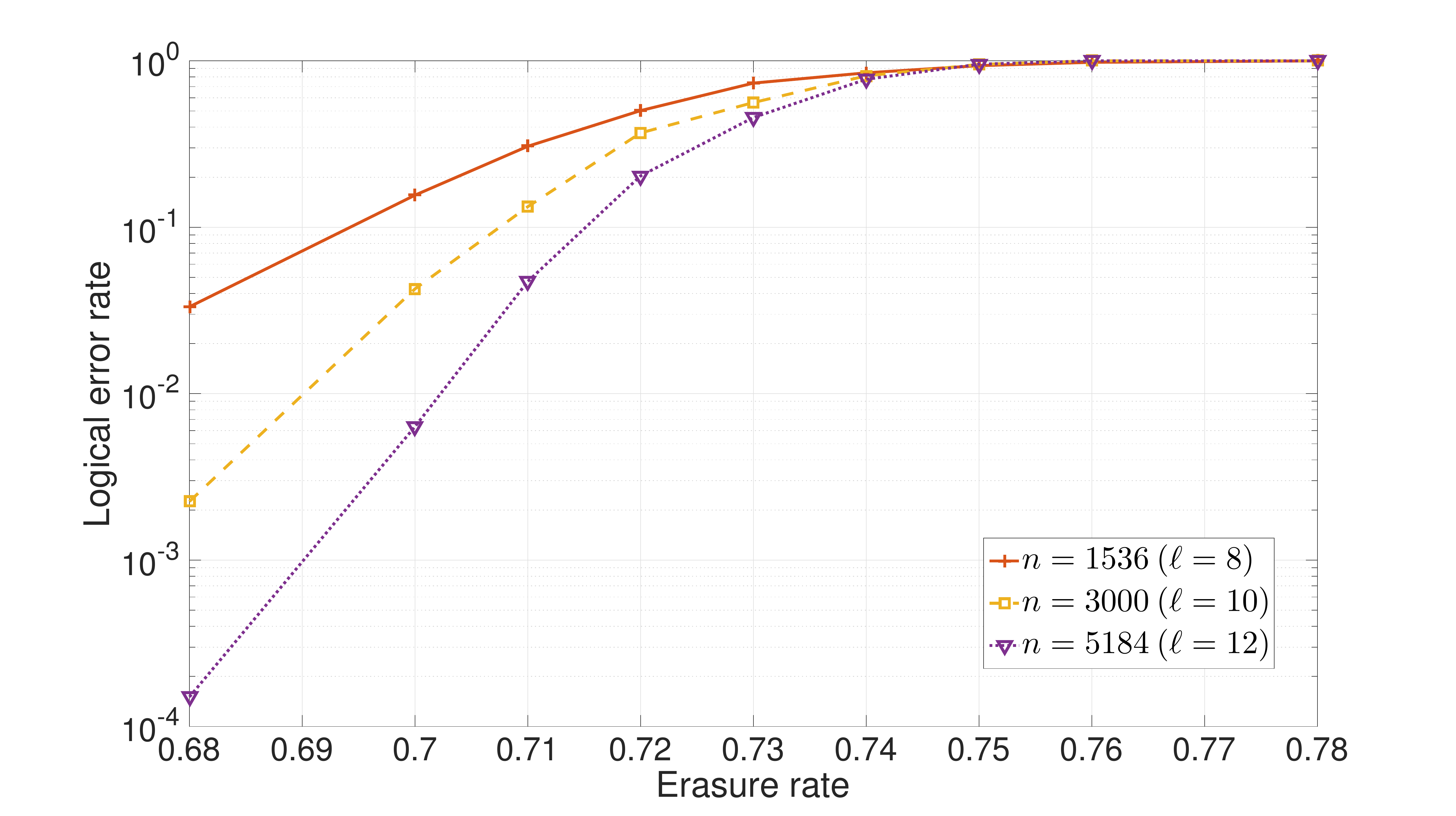}
\caption{Performance of Algorithm~\ref{alg:css-z-decoder} for erasure induced $X$ errors in 3D toric code with periodic boundary.}
\label{fig:x-sims-3dtc}
\end{figure}

The performance of the proposed algorithm for correcting the erasure induced $X$ errors is shown in Fig.~\ref{fig:x-sims-3dtc}.
We can see that the performance of the 3D toric code over the erasure channel is limited by its ability to correct phase errors. 

\begin{figure}[H]
\includegraphics[scale=0.17]{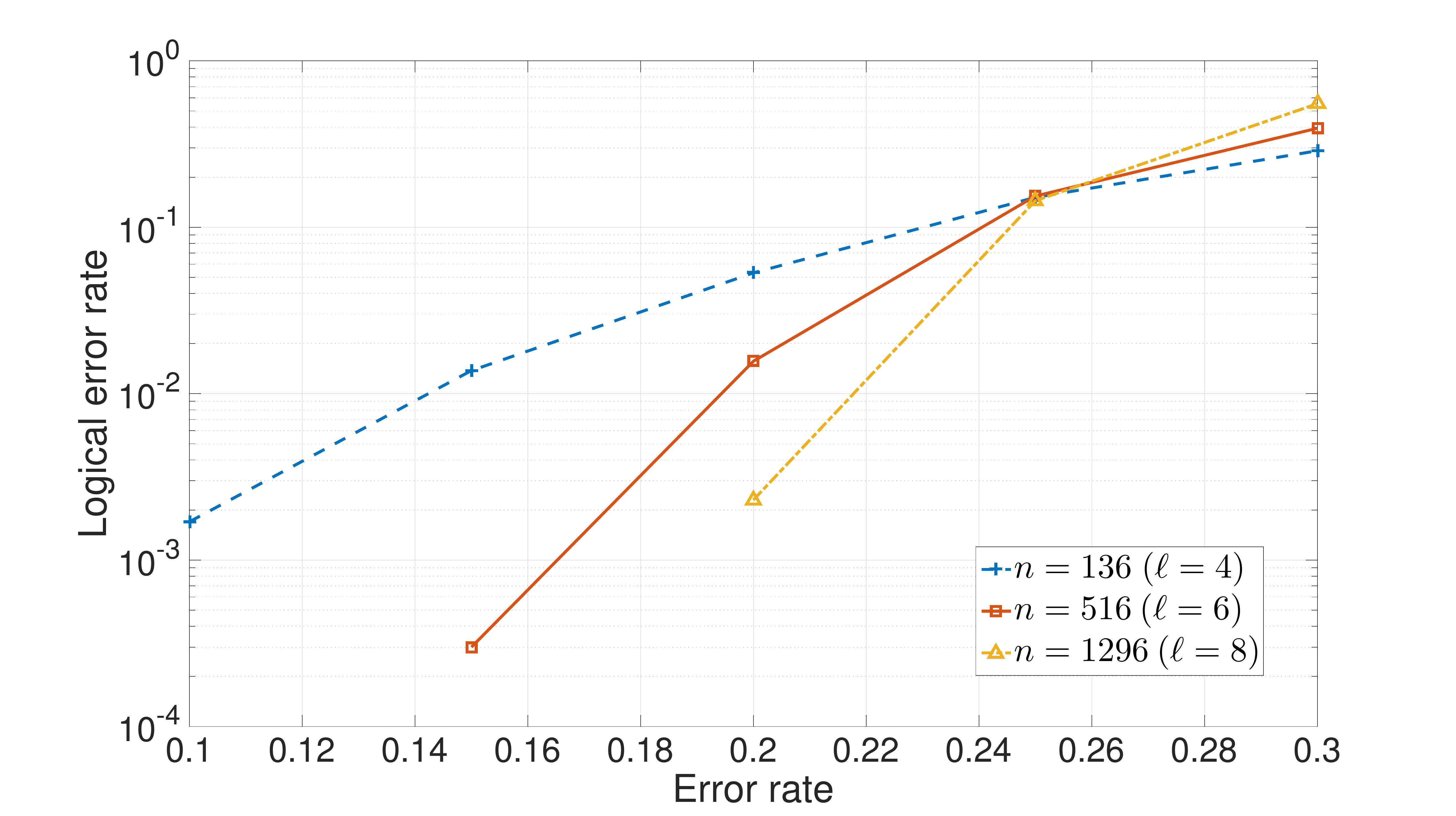}
\caption{Performance of correcting Z errors  under erasures in solid code.}
\label{fig:z-sims-solid-code}
\end{figure}

\begin{figure}[H]
\includegraphics[scale=0.17]{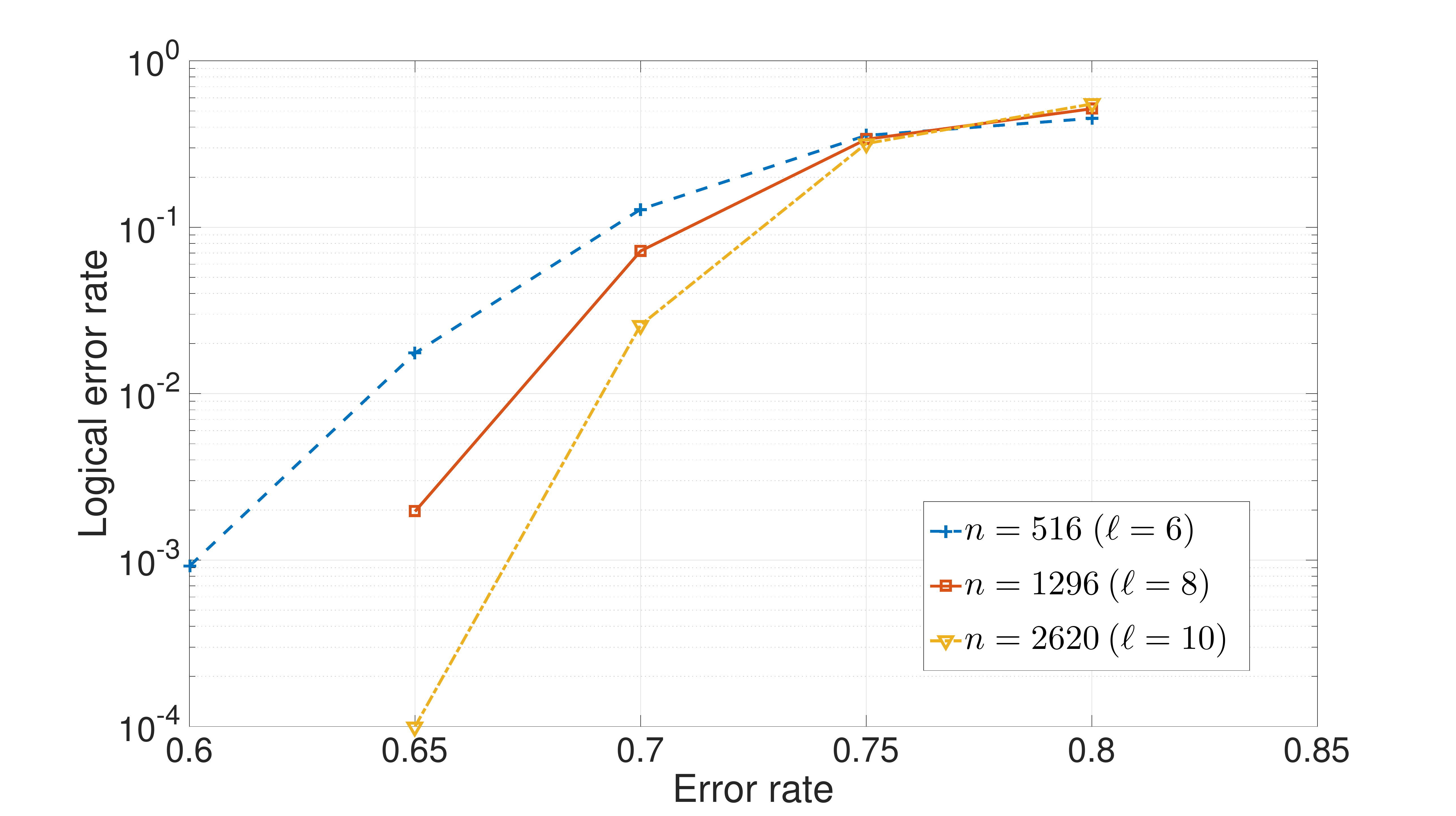}
\caption{Performance of correcting X errors  under erasures in solid code.}
\label{fig:x-sims-solid-code}
\end{figure}

Fig.~\ref{fig:z-sims-solid-code} shows the performance of   solid code for  for erasure induced phase flip errors. 
Fig.~\ref{fig:x-sims-solid-code} shows performance of  solid code for erasure induced bit flip errors.

\section{Decoding Welded Codes over the Erasure Channel}\label{sec:welded} 

In this section we propose a decoder for the welded code over the quantum erasure channel. 
Since the welded code is built from the 3D toric code, one might expect that the decoding of welded codes could be reduced to that of 3D toric codes. 
This is not exactly the case because the various component 3D toric codes are not entirely independent.  
The welded code is asymmetric and
as in the case of the 3D toric code, we need two different decoders for bit flip and phase flip errors.

We restrict our attention to the quantum erasure channel. 
{In correction of both phase and bit flip errors we  focus on unwelded erased qubits first and take decisions on them.
 Next we focus on the residual erased welded qubits and erased unwelded qubits on which decision was not made. }
In Section~\ref{ssec:welded-z-err}, we discuss decoder for correction of  
erasure induced phase errors.
And in section, \ref{ssec:welded-x-err} we describe the decoding algorithm for correction of  erasure induced bit flip errors. 

\subsection{Decoder for Z errors under erasure channel}\label{ssec:welded-z-err}

\begin{figure}[H]
    \centering
    \includegraphics[scale=2]{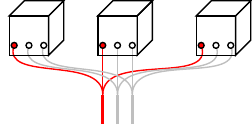}
    \caption{
Error on a single welded qubit  causes multiple nonzero syndromes unlike qubits in the 3D toric code which cause only two nonzero syndromes.}
    \label{fig:WeldedCodeZErrors2}
\end{figure}

In this section we will decode phase errors (induced by the erasures) on welded code based on the phase error decoder for 3d toric code.
Before that, let us see how phase errors in welded code differ phase errors in the solid code.
In the 3D toric code each qubit participates in exactly two $X$-type checks. 
Therefore, a single phase error flips exactly two qubits 
and one check if there are boundaries. 
In the welded code, a single phase error on a welded qubit can flip $O(R^2)$ $X$-type checks. 
 Fig.~\ref{fig:WeldedCodeZErrors2} illustrates the effect of a phase error on a welded qubit.

From a graphical point of view, 
welded qubits are no longer living on edges but hyperedges. 
A hyperedge is an edge that is incident on more than two  vertices. 
A phase error on a non-welded qubit behaves similar to a phase error on the toric code. 
It cause exactly two nonzero syndromes. 

Now let us look at the $Z$ stabilizers. 
Some of them are the stabilizers of the constituent toric codes.
These do not involve welded qubits. 
They continue to be cycles in the lattice on which the welded code is defined. 
The structure of stabilizers which involve welded qubits is slightly different. 
These welded $Z$-stabilizers are no longer cycles but hypercycles.
(A hypercycle $\sigma$ is a collection of edges such that every vertex has even degree with respect to the edges in $\sigma$.)

At this point we can use Algorithm~\ref{alg:css-z-decoder}. The important difference with respect to the 3D toric code comes in the step which requires the identification of qubits which can be frozen. 
From the previous discussion it follows that we need to identify hypercycles in the welded lattice. 
Every hypercycle will give us one qubit to be frozen.
At the end we will be left with a lattice without hypercycles. 
In other words, our goal is to find the spanning forest of the welded lattice. 
Unfortunately, finding  a forest in the welded lattice seems to be a hard problem.

We propose the following approach. 
Suppose that there are no erasures on the welded qubits. 
Then the decoding problem reduces to decoding a collection of independent 3D toric codes. 
In this case we could simply decode the various component toric codes and the combine the individual estimates. 
We can use the Algorithm~\ref{alg:css-z-decoder}.

However, if there are erasures on the welded qubits, then we cannot proceed in this fashion. 
We try to induce this situation by unerasing the welded qubits. 

We decode as many erasures as possible using peeling. 
Once peeling cannot proceed further, we have to identify the qubits which can be frozen. 
Recall that these come from the support of the stabilizers in the erased qubits. 
Due to the fact stabilizers no longer correspond to cycles in the  lattice of the welded code, we cannot simply find the spanning forest of the lattice of erased qubits. 
So, we first remove the welded qubits from the equation. 
Then we can identify the qubits which are part of stabilizers which are cycles. 
From each independent cycle we  obtain a qubit which can be frozen. 
We freeze them and reintroduce the welded qubits that were unerased and try to peel the associated Tanner graph. 

If the syndromes are cleared then we have been able to solve the decoding problem. 
On the other hand, if the syndromes cannot be cleared by peeling,
we have to use alternate methods to estimate the error on the residual erased qubits. 
Now at this point if any syndromes are left, it is because of stabilizers in support of erased qubits which are completely or partially on welded qubits. To correct such patterns we solve system of linear equations using Gaussian elimination. 
We summarize this procedure in 
Algorithm~\ref{alg:Welded decoder alg for Z errors}.

\begin{algorithm}[H]
\caption{Decoding $Z$ errors on erased qubits of welded code}
\begin{algorithmic}[1]
\REQUIRE {Erasure set $\mathcal{E}, $ $X$-syndromes $\sigma$,  and check matrix  $H$ }

\ENSURE {Error estimate consistent $f$ with measured syndrome ie $H_{\mathcal{E}} f^t = \sigma$}
\STATE $\mathcal{A} = \mathcal{E}$ \COMMENT $\mathcal{A}$ keeps track of currently unresolved qubit erasures.
\STATE Construct  Tanner graph, $\mathcal{T_A}$ defined by $H_{\mathcal{A}}$ 
\STATE Peel $\mathcal{T_A}$ using Algorithm~\ref{alg:peeling}. \COMMENT This updates the set $\mathcal{A}$.

\IF{nonzero syndromes exist}

\STATE Let $\mathcal{E}_w\subseteq \mathcal{A}$ be the set of erased welded qubits which have not been resolved yet. 
\STATE Let $\mathcal{B} = \mathcal{A}\setminus \mathcal{E}_w$ \COMMENT Remove the erased welded  qubits.
\STATE Find a spanning forest $\mathcal{F}$ in the erased lattice consisting of qubits in  $\mathcal{B}$
\STATE Freeze qubits in $\mathcal{B} \setminus \mathcal{F}$ \COMMENT These are nonwelded qubits not in the forest
\STATE Update $\mathcal{A} = \mathcal{F}\cup \mathcal{E}_w$
\COMMENT Reintroduce  erased welded  qubits.
\STATE Peel $\mathcal{T_A},\syn(\mathcal{A})$ using Algorithm~\ref{alg:peeling} \COMMENT Peeling updates the erasure set $\mathcal{A}$ and $\syn(\mathcal{A})$
\IF {$\syn(\mathcal{A})\neq0$ after peeling}
\STATE Solve  the system of equations 
$H_{\mathcal{A}} e^t = \syn(\mathcal{A})$ (using Gaussian elimination)
\ENDIF
\ENDIF
\end{algorithmic}
\label{alg:Welded decoder alg for Z errors}
\end{algorithm}

Note that in this process we do not peel until the welded qubits are reintroduced. 
The removal of the welded qubits is just to find the stabilizers in the support of the interior qubits ie the non-welded qubits.

\subsection{Decoder for X errors on erased qubits} \label{ssec:welded-x-err}

Let us now turn our attention to the $X$-errors.
We observed that the qubits live on the welded edges which can be viewed as hyperedges. 
The lattice for the welded code is  a hypergraph. 

In case of 3D toric code, we noted that the dual lattice is much more convenient to work with in the context of decoding $X$ errors. 
However, defining the dual lattice of the welded lattice is somewhat complicated and technical.
Instead, let us view the welded code as being constructed from the 3D toric codes which are represented in the dual lattice. 
Welding then leads to identification of the faces. 
Welding causes the identification of faces of different copies of the 3D toric code.
In effect this creates hyperfaces which can be incident on more than two volumes. 

Since the welded codes under consideration are obtained by the 
rough weld, $X$-stabilizers remain the same. 
However, because of the fact that faces can now be in the boundary of more than two volumes,
we lose the topological interpretation of $X$-stabilizers being the boundary of a closed volume. 
This is particularly true for an $X$-type stabilizer which has welded qubits in its support. 

This implies that the trapping algorithm used for the 3D toric codes to find the stabilizers in the support of erased qubits cannot be used. 
So we take an approach that is similar to the decoding of 
$Z$ errors. 

First, we apply the peeling algorithm to clear as many erasures as possible. 
Once the peeling algorithm gets stuck, we first identify an $X$-stabilizer which has no support on the welded qubits. 
This can be achieved by means of the trapping algorithm. 
We freeze one distinct qubit of the stabilizer and then apply peeling again. 
We repeat this process until we no longer find any more stabilizers in the set of erased qubits that are not welded.
If the syndrome has not been cleared, we  solve the system of linear equations using  Gaussian elimination. 
In our implementation the 
 complete algorithm uses  Algorithm~\ref{alg:dec-z-alternating} where we alternate between freezing and peeling.

\begin{algorithm}[H]
\caption{Decoder for X errors on erased qubits of welded code}
\begin{algorithmic}[1]
\REQUIRE {Erased qubits $\mathcal{E}$, $Z$-syndromes  $\tau$,  and check matrix $T$} 
\ENSURE {Error estimate $e$  such that  $T_{\mathcal{E}}e^t=\tau$
}
\STATE Initialize $\mathcal{A}= \mathcal{E}$
\STATE Construct Tanner graph $\mathcal{T_A}$ defined by restricted check matrix $T_{\mathcal{A}}$   

\STATE Peel $\mathcal{T_A}$ using Algorithm
~\ref{alg:peeling}. \COMMENT This updates the set $\mathcal{A}$ and the estimate $e$.
\WHILE {$\syn(\mathcal{A})\neq 0$}

\STATE Let $\mathcal{E}_w\subseteq \mathcal{A}$ be the set of erased welded qubits which have not been resolved. 
\STATE Let $\mathcal{B} = \mathcal{A}\setminus \mathcal{E}_w$ \COMMENT Unerase the erased welded  qubits.
\STATE Call trapping algorithm for each solid code separately and find a stabilizer with support in $\mathcal{B}$. 
\STATE In $\mathcal{T_A}$ randomly freeze one qubit in the support of the stabilizer  obtained in previous step.
\STATE Peel $\mathcal{T_A}$ \COMMENT Updates the error estimate $e$
\ENDWHILE

\IF {nonzero syndromes remain} \COMMENT Solve for unresolved errors in $e$
\STATE Solve system of linear equations by Gaussian elimination. 
\ENDIF
\STATE Return the estimate $e$.
\end{algorithmic}
\label{alg:Welded decoder alg for X errors}
\end{algorithm}

\subsection{Simulation results}
\begin{figure}[H]
    \centering
    \includegraphics[scale=0.16]{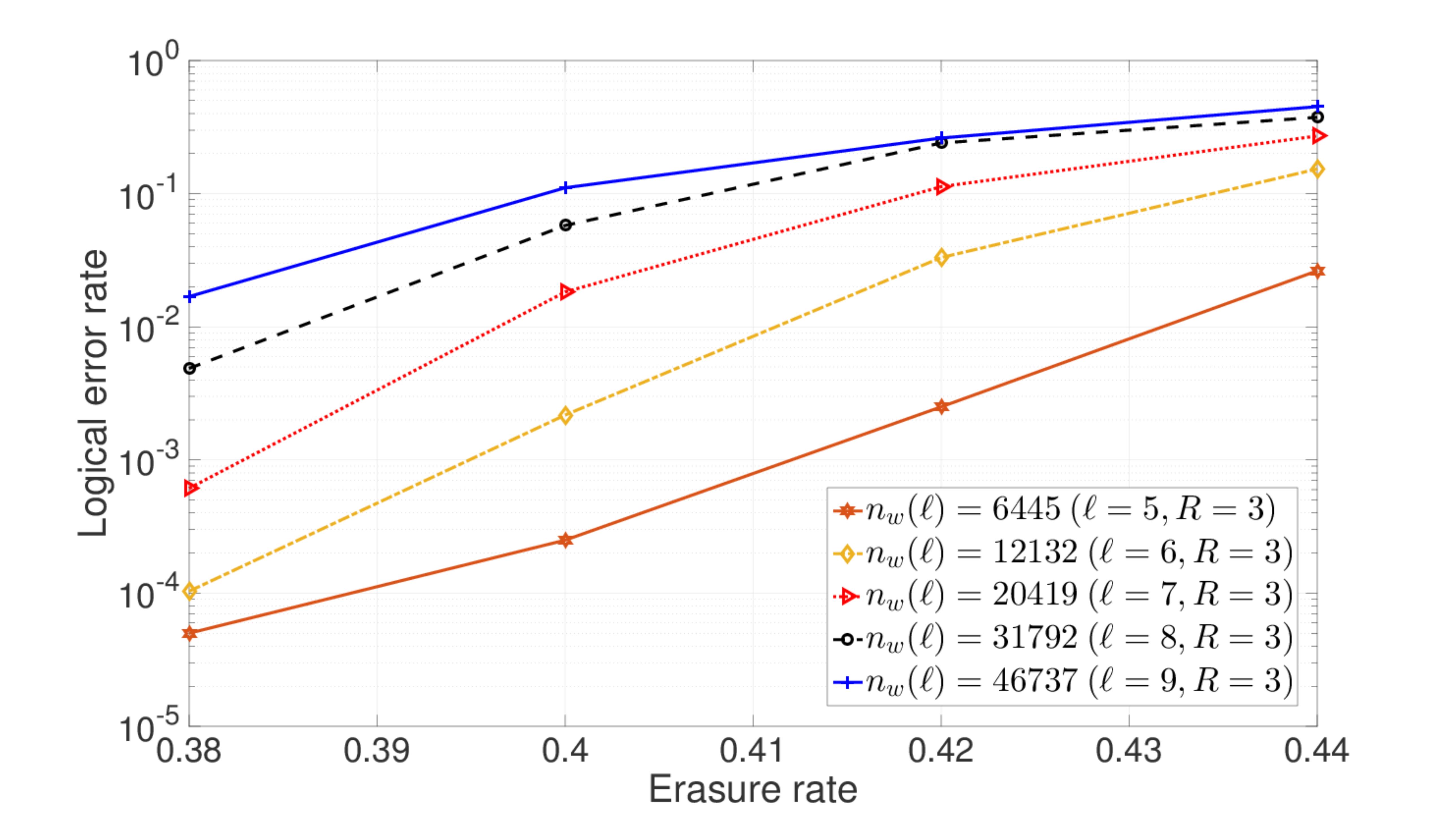}
    \caption{Welded code performance for erasure channel, for various values of $\ell$ and $R$.}
    \label{fig:WeldedCodePerfForErasure}
\end{figure}

\begin{figure}[H]
    \centering
    \includegraphics[width=8cm,height=4.7cm]{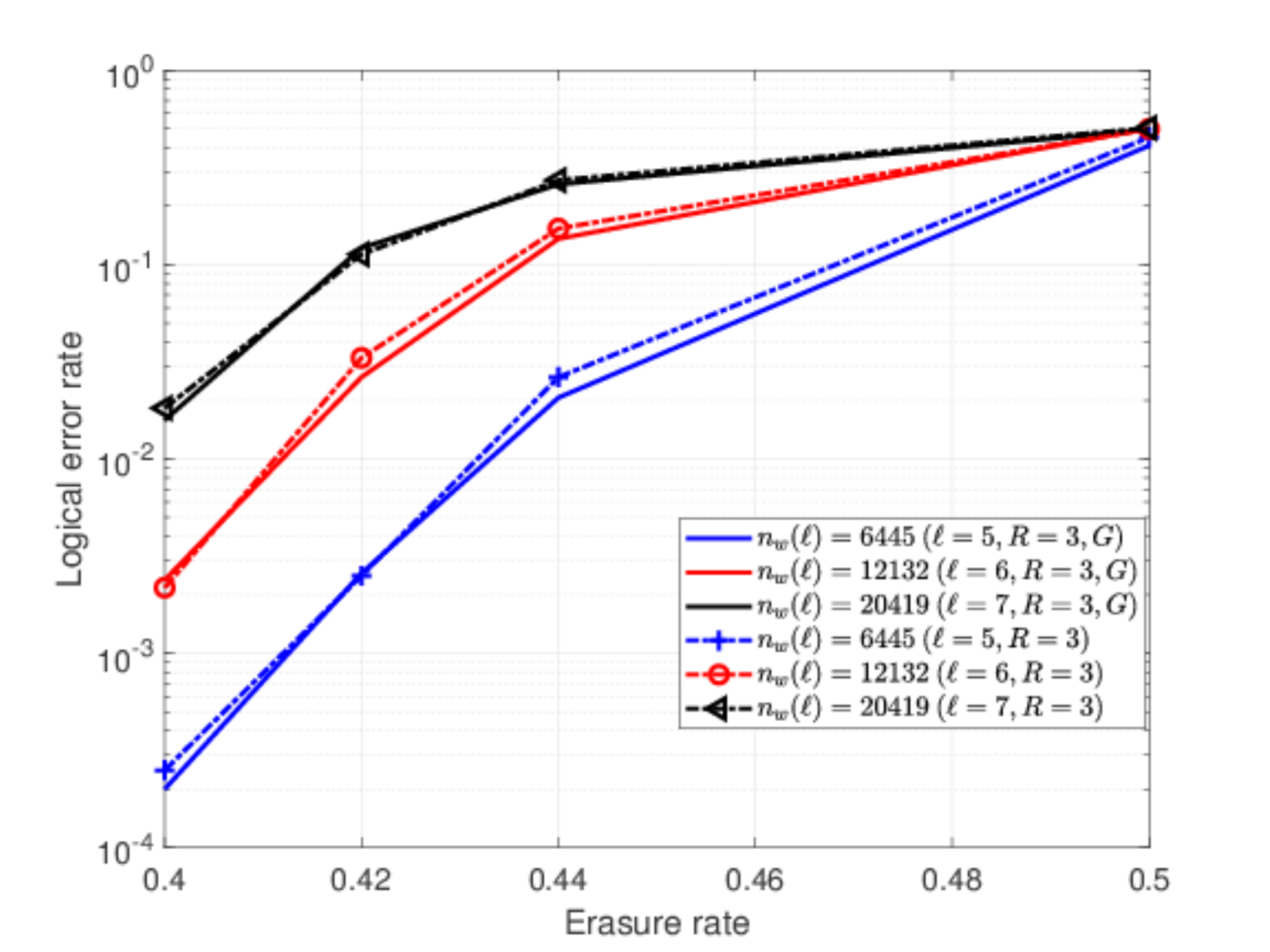}
    \caption{Welded code performance comparison for the proposed decoder (dashed lines)  and  one based on Gaussian elimination. The same input is given to both the decoders. In the legend, `\textit{G}'  denotes the decoder that solves system of linear equations using Gaussian elimination.}
    \label{fig:WeldedCodePerfForErasureGaussian}
\end{figure}

Here we present the performance to welded code for erasure decoder presented previously. 
As can be seen from  Fig.~\ref{fig:WeldedCodePerfForErasure}, no threshold is observed. 
To confirm that this is not an artifact of the proposed decoder, we also studied the performance of the welded codes
using a decoder that is based on Gaussian elimination. Here, we simply solve the system of equations that arise in the context of 
quantum erasure channel. 
We simulated both the decoders under the same settings i.e. same erasure patterns were given to both the decoders.  
Our results, see Fig.~\ref{fig:WeldedCodePerfForErasureGaussian}, show that there is not much difference between the proposed decoder and the decoder using Gaussian elimination. 

\section{Conclusion}\label{sec:conc}

In this paper we  proposed decoders for the 3D toric code on the cubic lattice with and without boundaries. 
We also studied the performance of these decoders numerically.
The proposed decoder for the 3D toric codes over the bit flip channel can be improved and generalized to  3D toric codes
on arbitrary lattices. 
We also reported the performance of the 3D toric code on the quantum erasure channel. 
Considering the observed threshold is very close to the bond percolation threshold of the cubic lattice, we expect the proposed erasure decoder to be almost optimal.

We also proposed an efficient decoder for the welded codes over the erasure channel. 
The results on welded codes prompt a closer look at the relation between the energy gap of the code and the code threshold. 
The toric codes have a constant energy barrier but a high threshold. 
The cubic code which has logarithmic energy gap has a lower threshold of about 2\%. 
The welded codes have the highest known energy gap but no threshold. 
One is tempted to conjecture that there might be a tradeoff between the threshold and the energy gap.
Understanding the relation  between threshold and energy barrier  would be an interesting problem for future research. 
Another interesting direction would be to come up with new constructions of  codes that have a high threshold and also a nonconstant  
energy gap.


\begin{thebibliography}{24}%
\makeatletter
\providecommand \@ifxundefined [1]{%
 \@ifx{#1\undefined}
}%
\providecommand \@ifnum [1]{%
 \ifnum #1\expandafter \@firstoftwo
 \else \expandafter \@secondoftwo
 \fi
}%
\providecommand \@ifx [1]{%
 \ifx #1\expandafter \@firstoftwo
 \else \expandafter \@secondoftwo
 \fi
}%
\providecommand \natexlab [1]{#1}%
\providecommand \enquote  [1]{``#1''}%
\providecommand \bibnamefont  [1]{#1}%
\providecommand \bibfnamefont [1]{#1}%
\providecommand \citenamefont [1]{#1}%
\providecommand \href@noop [0]{\@secondoftwo}%
\providecommand \href [0]{\begingroup \@sanitize@url \@href}%
\providecommand \@href[1]{\@@startlink{#1}\@@href}%
\providecommand \@@href[1]{\endgroup#1\@@endlink}%
\providecommand \@sanitize@url [0]{\catcode `\\12\catcode `\$12\catcode
  `\&12\catcode `\#12\catcode `\^12\catcode `\_12\catcode `\%12\relax}%
\providecommand \@@startlink[1]{}%
\providecommand \@@endlink[0]{}%
\providecommand \url  [0]{\begingroup\@sanitize@url \@url }%
\providecommand \@url [1]{\endgroup\@href {#1}{\urlprefix }}%
\providecommand \urlprefix  [0]{URL }%
\providecommand \Eprint [0]{\href }%
\providecommand \doibase [0]{http://dx.doi.org/}%
\providecommand \selectlanguage [0]{\@gobble}%
\providecommand \bibinfo  [0]{\@secondoftwo}%
\providecommand \bibfield  [0]{\@secondoftwo}%
\providecommand \translation [1]{[#1]}%
\providecommand \BibitemOpen [0]{}%
\providecommand \bibitemStop [0]{}%
\providecommand \bibitemNoStop [0]{.\EOS\space}%
\providecommand \EOS [0]{\spacefactor3000\relax}%
\providecommand \BibitemShut  [1]{\csname bibitem#1\endcsname}%
\let\auto@bib@innerbib\@empty
\bibitem [{\citenamefont {Kubica}\ \emph {et~al.}(2015)\citenamefont {Kubica},
  \citenamefont {Yoshida},\ and\ \citenamefont {Pastawski}}]{kubica15}%
  \BibitemOpen
  \bibfield  {author} {\bibinfo {author} {\bibfnamefont {A.}~\bibnamefont
  {Kubica}}, \bibinfo {author} {\bibfnamefont {B.}~\bibnamefont {Yoshida}}, \
  and\ \bibinfo {author} {\bibfnamefont {F.}~\bibnamefont {Pastawski}},\ }\href
  {http://stacks.iop.org/1367-2630/17/i=8/a=083026} {\bibfield  {journal}
  {\bibinfo  {journal} {New Journal of Physics}\ }\textbf {\bibinfo {volume}
  {17}},\ \bibinfo {pages} {083026} (\bibinfo {year} {2015})}\BibitemShut
  {NoStop}%
\bibitem [{\citenamefont {Aloshious}\ and\ \citenamefont
  {Sarvepalli}(2016)}]{aloshious16}%
  \BibitemOpen
  \bibfield  {author} {\bibinfo {author} {\bibfnamefont {A.~B.}\ \bibnamefont
  {Aloshious}}\ and\ \bibinfo {author} {\bibfnamefont {P.~K.}\ \bibnamefont
  {Sarvepalli}},\ }\href {https://arxiv.org/abs/1606.00960} {\enquote {\bibinfo
  {title} {Projecting {3D} color codes onto {3D} toric codes},}\ } (\bibinfo
  {year} {2016}),\ \bibinfo {note} {arXiv:1606.00960}\BibitemShut {NoStop}%
\bibitem [{\citenamefont {Michnicki}(2014)}]{michnicki14}%
  \BibitemOpen
  \bibfield  {author} {\bibinfo {author} {\bibfnamefont {K.~P.}\ \bibnamefont
  {Michnicki}},\ }\href {\doibase 10.1103/PhysRevLett.113.130501} {\bibfield
  {journal} {\bibinfo  {journal} {Phys. Rev. Lett.}\ }\textbf {\bibinfo
  {volume} {113}},\ \bibinfo {pages} {130501} (\bibinfo {year}
  {2014})}\BibitemShut {NoStop}%
\bibitem [{\citenamefont {Siva}\ and\ \citenamefont {Yoshida}(2017)}]{siva17}%
  \BibitemOpen
  \bibfield  {author} {\bibinfo {author} {\bibfnamefont {K.}~\bibnamefont
  {Siva}}\ and\ \bibinfo {author} {\bibfnamefont {B.}~\bibnamefont {Yoshida}},\
  }\href {\doibase 10.1103/PhysRevA.95.032324} {\bibfield  {journal} {\bibinfo
  {journal} {Phys. Rev. A}\ }\textbf {\bibinfo {volume} {95}},\ \bibinfo
  {pages} {032324} (\bibinfo {year} {2017})}\BibitemShut {NoStop}%
\bibitem [{\citenamefont {Ioffe}\ and\ \citenamefont
  {M\'ezard}(2007)}]{ioffe08}%
  \BibitemOpen
  \bibfield  {author} {\bibinfo {author} {\bibfnamefont {L.}~\bibnamefont
  {Ioffe}}\ and\ \bibinfo {author} {\bibfnamefont {M.}~\bibnamefont
  {M\'ezard}},\ }\href {\doibase 10.1103/PhysRevA.75.032345} {\bibfield
  {journal} {\bibinfo  {journal} {Phys. Rev. A}\ }\textbf {\bibinfo {volume}
  {75}},\ \bibinfo {pages} {032345} (\bibinfo {year} {2007})}\BibitemShut
  {NoStop}%
\bibitem [{\citenamefont {Brooks}\ and\ \citenamefont
  {Preskill}(2013)}]{brooks13}%
  \BibitemOpen
  \bibfield  {author} {\bibinfo {author} {\bibfnamefont {P.}~\bibnamefont
  {Brooks}}\ and\ \bibinfo {author} {\bibfnamefont {J.}~\bibnamefont
  {Preskill}},\ }\href {\doibase 10.1103/PhysRevA.87.032310} {\bibfield
  {journal} {\bibinfo  {journal} {Phys. Rev. A}\ }\textbf {\bibinfo {volume}
  {87}},\ \bibinfo {pages} {032310} (\bibinfo {year} {2013})}\BibitemShut
  {NoStop}%
\bibitem [{\citenamefont {Dennis}\ \emph {et~al.}(2002)\citenamefont {Dennis},
  \citenamefont {Kitaev}, \citenamefont {Landahl},\ and\ \citenamefont
  {Preskill}}]{dennis02}%
  \BibitemOpen
  \bibfield  {author} {\bibinfo {author} {\bibfnamefont {E.}~\bibnamefont
  {Dennis}}, \bibinfo {author} {\bibfnamefont {A.}~\bibnamefont {Kitaev}},
  \bibinfo {author} {\bibfnamefont {A.}~\bibnamefont {Landahl}}, \ and\
  \bibinfo {author} {\bibfnamefont {J.}~\bibnamefont {Preskill}},\ }\href@noop
  {} {\bibfield  {journal} {\bibinfo  {journal} {J. Math. Phys. 43, 4452-4505}\
  } (\bibinfo {year} {2002})}\BibitemShut {NoStop}%
\bibitem [{\citenamefont {Wang}\ \emph {et~al.}(2010)\citenamefont {Wang},
  \citenamefont {Fowler}, \citenamefont {Stephens},\ and\ \citenamefont
  {Hollenberg}}]{wang10}%
  \BibitemOpen
  \bibfield  {author} {\bibinfo {author} {\bibfnamefont {D.~S.}\ \bibnamefont
  {Wang}}, \bibinfo {author} {\bibfnamefont {A.~G.}\ \bibnamefont {Fowler}},
  \bibinfo {author} {\bibfnamefont {A.~M.}\ \bibnamefont {Stephens}}, \ and\
  \bibinfo {author} {\bibfnamefont {L.~C.~L.}\ \bibnamefont {Hollenberg}},\
  }\href@noop {} {\bibfield  {journal} {\bibinfo  {journal} {Quant. Inf.
  Comput.}\ }\textbf {\bibinfo {volume} {10}},\ \bibinfo {pages} {456}
  (\bibinfo {year} {2010})}\BibitemShut {NoStop}%
\bibitem [{\citenamefont {Duivenvoorden}\ \emph {et~al.}(2017)\citenamefont
  {Duivenvoorden}, \citenamefont {Breuckmann},\ and\ \citenamefont
  {Terhal}}]{duivenvoorden17}%
  \BibitemOpen
  \bibfield  {author} {\bibinfo {author} {\bibfnamefont {K.}~\bibnamefont
  {Duivenvoorden}}, \bibinfo {author} {\bibfnamefont {N.~P.}\ \bibnamefont
  {Breuckmann}}, \ and\ \bibinfo {author} {\bibfnamefont {B.~M.}\ \bibnamefont
  {Terhal}},\ }\href {https://arxiv.org/abs/1708.09286} {\enquote {\bibinfo
  {title} {Renormalization group decoder for a four-dimensional toric code},}\
  } (\bibinfo {year} {2017}),\ \bibinfo {note} {arXiv:1708.09286}\BibitemShut
  {NoStop}%
\bibitem [{\citenamefont {Delfosse}\ and\ \citenamefont
  {Z{\'{e}}mor}(2017)}]{delfosse17}%
  \BibitemOpen
  \bibfield  {author} {\bibinfo {author} {\bibfnamefont {N.}~\bibnamefont
  {Delfosse}}\ and\ \bibinfo {author} {\bibfnamefont {G.}~\bibnamefont
  {Z{\'{e}}mor}},\ }\href {http://arxiv.org/abs/1703.01517} {\enquote {\bibinfo
  {title} {Linear-time maximum likelihood decoding of surface codes over the
  quantum erasure channel},}\ } (\bibinfo {year} {2017}),\ \bibinfo {note}
  {arXiv:1703.01517}\BibitemShut {NoStop}%
\bibitem [{\citenamefont {Wilke}(1983)}]{wilke83}%
  \BibitemOpen
  \bibfield  {author} {\bibinfo {author} {\bibfnamefont {S.}~\bibnamefont
  {Wilke}},\ }\href {https://doi.org/10.1016/0375-9601(83)90005-1} {\bibfield
  {journal} {\bibinfo  {journal} {Physics Letters A}\ }\textbf {\bibinfo
  {volume} {96}},\ \bibinfo {pages} {344} (\bibinfo {year} {1983})}\BibitemShut
  {NoStop}%
\bibitem [{\citenamefont {Michnicki}(2012)}]{arxiv_michnicki14}%
  \BibitemOpen
  \bibfield  {author} {\bibinfo {author} {\bibfnamefont {K.~P.}\ \bibnamefont
  {Michnicki}},\ }\href {https://arxiv.org/abs/1208.3496} {\enquote {\bibinfo
  {title} {{3D} topological quantum memory with a power-law energy barrier},}\
  } (\bibinfo {year} {2012}),\ \bibinfo {note} {arXiv:1208.3496}\BibitemShut
  {NoStop}%
\bibitem [{\citenamefont {Calderbank}\ \emph {et~al.}(1998)\citenamefont
  {Calderbank}, \citenamefont {Rains}, \citenamefont {Shor},\ and\
  \citenamefont {Sloane}}]{calderbank98}%
  \BibitemOpen
  \bibfield  {author} {\bibinfo {author} {\bibfnamefont {A.~R.}\ \bibnamefont
  {Calderbank}}, \bibinfo {author} {\bibfnamefont {E.~M.}\ \bibnamefont
  {Rains}}, \bibinfo {author} {\bibfnamefont {P.~M.}\ \bibnamefont {Shor}}, \
  and\ \bibinfo {author} {\bibfnamefont {N.~J.~A.}\ \bibnamefont {Sloane}},\
  }\href {\doibase 10.1109/18.681315} {\bibfield  {journal} {\bibinfo
  {journal} {IEEE Trans. on Inform. Theory}\ }\textbf {\bibinfo {volume}
  {44}},\ \bibinfo {pages} {1369} (\bibinfo {year} {1998})}\BibitemShut
  {NoStop}%
\bibitem [{\citenamefont {Pastawski}\ \emph {et~al.}(2011)\citenamefont
  {Pastawski}, \citenamefont {Clemente},\ and\ \citenamefont
  {Cirac}}]{pastawski11}%
  \BibitemOpen
  \bibfield  {author} {\bibinfo {author} {\bibfnamefont {F.}~\bibnamefont
  {Pastawski}}, \bibinfo {author} {\bibfnamefont {L.}~\bibnamefont {Clemente}},
  \ and\ \bibinfo {author} {\bibfnamefont {J.~I.}\ \bibnamefont {Cirac}},\
  }\href {\doibase 10.1103/PhysRevA.83.012304} {\bibfield  {journal} {\bibinfo
  {journal} {Phys. Rev. A}\ }\textbf {\bibinfo {volume} {83}},\ \bibinfo
  {pages} {012304} (\bibinfo {year} {2011})}\BibitemShut {NoStop}%
\bibitem [{\citenamefont {Breuckmann}\ \emph {et~al.}(2017)\citenamefont
  {Breuckmann}, \citenamefont {Duivenvoorden}, \citenamefont {Michels},\ and\
  \citenamefont {Terhal}}]{breuckmann16}%
  \BibitemOpen
  \bibfield  {author} {\bibinfo {author} {\bibfnamefont {N.}~\bibnamefont
  {Breuckmann}}, \bibinfo {author} {\bibfnamefont {K.}~\bibnamefont
  {Duivenvoorden}}, \bibinfo {author} {\bibfnamefont {D.}~\bibnamefont
  {Michels}}, \ and\ \bibinfo {author} {\bibfnamefont {B.}~\bibnamefont
  {Terhal}},\ }\href@noop {} {\bibfield  {journal} {\bibinfo  {journal}
  {Quantum Information \& Computation}\ }\textbf {\bibinfo {volume} {17}},\
  \bibinfo {pages} {181} (\bibinfo {year} {2017})}\BibitemShut {NoStop}%
\bibitem [{\citenamefont {Grassl}\ \emph {et~al.}(1997)\citenamefont {Grassl},
  \citenamefont {Beth},\ and\ \citenamefont {Pellizzari}}]{grassl97}%
  \BibitemOpen
  \bibfield  {author} {\bibinfo {author} {\bibfnamefont {M.}~\bibnamefont
  {Grassl}}, \bibinfo {author} {\bibfnamefont {T.}~\bibnamefont {Beth}}, \ and\
  \bibinfo {author} {\bibfnamefont {T.}~\bibnamefont {Pellizzari}},\ }\href
  {\doibase 10.1103/PhysRevA.56.33} {\bibfield  {journal} {\bibinfo  {journal}
  {Phys. Rev. A}\ }\textbf {\bibinfo {volume} {56}},\ \bibinfo {pages} {33}
  (\bibinfo {year} {1997})}\BibitemShut {NoStop}%
\bibitem [{\citenamefont {Kudekar}\ \emph {et~al.}(2016)\citenamefont
  {Kudekar}, \citenamefont {Kumar}, \citenamefont {Mondelli}, \citenamefont
  {Pfister}, \citenamefont {\c{S}a\c{s}o\u{g}lu},\ and\ \citenamefont
  {Urbanke}}]{kudekar16}%
  \BibitemOpen
  \bibfield  {author} {\bibinfo {author} {\bibfnamefont {S.}~\bibnamefont
  {Kudekar}}, \bibinfo {author} {\bibfnamefont {S.}~\bibnamefont {Kumar}},
  \bibinfo {author} {\bibfnamefont {M.}~\bibnamefont {Mondelli}}, \bibinfo
  {author} {\bibfnamefont {H.~D.}\ \bibnamefont {Pfister}}, \bibinfo {author}
  {\bibfnamefont {E.}~\bibnamefont {\c{S}a\c{s}o\u{g}lu}}, \ and\ \bibinfo
  {author} {\bibfnamefont {R.}~\bibnamefont {Urbanke}},\ }in\ \href
  {http://doi.acm.org/10.1145/2897518.2897584} {\emph {\bibinfo {booktitle}
  {Proceedings of the Forty-eighth Annual ACM Symposium on Theory of
  Computing}}},\ \bibinfo {series and number} {STOC '16}\ (\bibinfo {year}
  {2016})\ pp.\ \bibinfo {pages} {658--669}\BibitemShut {NoStop}%
\bibitem [{\citenamefont {Lloyd}\ \emph {et~al.}(2017)\citenamefont {Lloyd},
  \citenamefont {Shor},\ and\ \citenamefont {Thompson}}]{lloyd17}%
  \BibitemOpen
  \bibfield  {author} {\bibinfo {author} {\bibfnamefont {S.}~\bibnamefont
  {Lloyd}}, \bibinfo {author} {\bibfnamefont {P.}~\bibnamefont {Shor}}, \ and\
  \bibinfo {author} {\bibfnamefont {K.}~\bibnamefont {Thompson}},\ }\href
  {http://arxiv.org/abs/1703.00382} {\enquote {\bibinfo {title} {polylog-{LDPC}
  capacity achieving codes for the noisy quantum erasure channel},}\ }
  (\bibinfo {year} {2017}),\ \bibinfo {note} {arXiv:1703.00382}\BibitemShut
  {NoStop}%
\bibitem [{\citenamefont {Delfosse}\ \emph {et~al.}(2016)\citenamefont
  {Delfosse}, \citenamefont {Iyer},\ and\ \citenamefont {Poulin}}]{delfosse16}%
  \BibitemOpen
  \bibfield  {author} {\bibinfo {author} {\bibfnamefont {N.}~\bibnamefont
  {Delfosse}}, \bibinfo {author} {\bibfnamefont {P.}~\bibnamefont {Iyer}}, \
  and\ \bibinfo {author} {\bibfnamefont {D.}~\bibnamefont {Poulin}},\ }\href
  {http://arxiv.org/abs/1611.04256} {\enquote {\bibinfo {title} {A linear-time
  benchmarking tool for generalized surface codes},}\ } (\bibinfo {year}
  {2016}),\ \bibinfo {note} {arXiv:1611.04256}\BibitemShut {NoStop}%
\bibitem [{\citenamefont {Delfosse}\ and\ \citenamefont
  {Z{\'e}mor}(2013)}]{delfosse13}%
  \BibitemOpen
  \bibfield  {author} {\bibinfo {author} {\bibfnamefont {N.}~\bibnamefont
  {Delfosse}}\ and\ \bibinfo {author} {\bibfnamefont {G.}~\bibnamefont
  {Z{\'e}mor}},\ }\href {http://dl.acm.org/citation.cfm?id=2535680.2535684}
  {\bibfield  {journal} {\bibinfo  {journal} {Quantum Information \&
  Computation}\ }\textbf {\bibinfo {volume} {13}},\ \bibinfo {pages} {793}
  (\bibinfo {year} {2013})}\BibitemShut {NoStop}%
\bibitem [{\citenamefont {Gottesman}(1997)}]{gottesman97}%
  \BibitemOpen
  \bibfield  {author} {\bibinfo {author} {\bibfnamefont {D.}~\bibnamefont
  {Gottesman}},\ }\href {https://arxiv.org/abs/quant-ph/9705052} {\enquote
  {\bibinfo {title} {Stabilizer codes and quantum error correction},}\ }
  (\bibinfo {year} {1997}),\ \bibinfo {note} {{C}altech Ph. D. Thesis, arXiv:
  quant-ph/9705052}\BibitemShut {NoStop}%
\bibitem [{\citenamefont {Castelnovo}\ and\ \citenamefont
  {Chamon}(2008)}]{castelnovo08}%
  \BibitemOpen
  \bibfield  {author} {\bibinfo {author} {\bibfnamefont {C.}~\bibnamefont
  {Castelnovo}}\ and\ \bibinfo {author} {\bibfnamefont {C.}~\bibnamefont
  {Chamon}},\ }\href {\doibase 10.1103/PhysRevB.78.155120} {\bibfield
  {journal} {\bibinfo  {journal} {Phys. Rev. B}\ }\textbf {\bibinfo {volume}
  {78}},\ \bibinfo {pages} {155120} (\bibinfo {year} {2008})}\BibitemShut
  {NoStop}%
\bibitem [{\citenamefont {Ohno}\ \emph {et~al.}(2004)\citenamefont {Ohno},
  \citenamefont {G.}, \citenamefont {Ikuo~Ichinose},\ and\ \citenamefont
  {Matsui}}]{ohno04}%
  \BibitemOpen
  \bibfield  {author} {\bibinfo {author} {\bibfnamefont {T.}~\bibnamefont
  {Ohno}}, \bibinfo {author} {\bibfnamefont {A.}~\bibnamefont {G.}}, \bibinfo
  {author} {\bibfnamefont {I.}~\bibnamefont {Ikuo~Ichinose}}, \ and\ \bibinfo
  {author} {\bibfnamefont {T.}~\bibnamefont {Matsui}},\ }\href
  {https://doi.org/10.1016/j.nuclphysb.2004.07.003} {\bibfield  {journal}
  {\bibinfo  {journal} {Nuclear Physics B}\ }\textbf {\bibinfo {volume}
  {697}},\ \bibinfo {pages} {462} (\bibinfo {year} {2004})}\BibitemShut
  {NoStop}%
\bibitem [{\citenamefont {Takeda}\ and\ \citenamefont
  {Nishimori}(2005)}]{takeda05}%
  \BibitemOpen
  \bibfield  {author} {\bibinfo {author} {\bibfnamefont {K.}~\bibnamefont
  {Takeda}}\ and\ \bibinfo {author} {\bibfnamefont {H.}~\bibnamefont
  {Nishimori}},\ }\href {https://doi.org/10.1143/JPSJS.74S.115} {\bibfield
  {journal} {\bibinfo  {journal} {Journal of the Physical Society of Japan}\
  }\textbf {\bibinfo {volume} {74}},\ \bibinfo {pages} {115} (\bibinfo {year}
  {2005})}\BibitemShut {NoStop}%
\end{thebibliography}
%

\end{document}